\documentclass[pdflatex,sn-vancouver,Numbered]{sn-jnl}

\usepackage{graphicx}%
\usepackage{multirow}%
\usepackage{amsmath,amssymb,amsfonts}%
\usepackage{amsthm}%
\usepackage{mathrsfs}%
\usepackage[title]{appendix}%
\usepackage{xcolor}%
\usepackage{textcomp}%
\usepackage{manyfoot}%
\usepackage{booktabs}%
\usepackage{algorithm}%
\usepackage{algorithmicx}%
\usepackage{algpseudocode}%
\usepackage{listings}%
\usepackage{subfig}%
\usepackage{aas_macros}%
\usepackage{lineno}%
\usepackage{ulem}%
\usepackage{float}%

\theoremstyle{thmstyleone}%
\theoremstyle{thmstyletwo}%

\theoremstyle{thmstylethree}%

\begin{document}

\title[Article Title]{Blazars Jets and prospects for TeV-PeV neutrinos $\&$ gamma-rays through cosmic-ray interactions}


\author*[1]{\fnm{Rodrigo} \sur{Sasse}}\email{rodrigo.sasse1@uel.br}

\author[2]{\fnm{Rubens Jr.} \sur{Costa}}\email{rubensp@utfpr.edu.br}
\equalcont{These authors contributed equally to this work.}

\author[3,4]{\fnm{Luiz A.} \sur{Stuani Pereira}}\email{luizstuani@uaf.ufcg.edu.br}
\equalcont{These authors contributed equally to this work.}

\author[1,2,5,6,7,8]{\fnm{Rita C.} \sur{dos Anjos}}\email{ritacassia@ufpr.br}
\equalcont{These authors contributed equally to this work.}

\affil*[1]{\orgdiv{Programa de p\'os-graduação em F\'isica \& Departamento de F\'isica}, \orgname{Universidade Estadual de Londrina (UEL)}, \orgaddress{\street{Rodovia Celso Garcia Cid Km 380}, \city{Londrina}, \postcode{86057-970}, \state{PR}, \country{Brazil}}}

\affil[2]{\orgdiv{Programa de Pós-Graduação em Física e Astronomia}, \orgname{Universidade Tecnológica Federal do Paraná (UTFPR)}, \orgaddress{\street{Av. Sete de Setembro, 3165}, \city{Curitiba}, \postcode{80230-901}, \state{PR}, \country{Brazil}}}

\affil[3]{\orgdiv{Unidade Acadêmica de Física}, \orgname{Universidade Federal de Campina Grande (UFCG)}, \orgaddress{\street{Rua Aprígio Veloso, 882}, \city{Campina Grande}, \postcode{58429-900}, \state{Para\'iba}, \country{Brazil}}}

\affil[4]{\orgdiv{Instituto de Física}, \orgname{Universidade de São Paulo (USP)}, \orgaddress{\street{Rua do Matão, 1371}, \city{São Paulo}, \postcode{05508-090}, \state{São Paulo}, \country{Brazil}}}

\affil[5]{\orgname{Max-Planck-Institut für Kernphysik}, \orgaddress{\street{Saupfercheckweg 1}, \city{Heidelberg}, \postcode{69117}, \country{Germany}}}

\affil[6]{\orgdiv{Departamento de Engenharias e Exatas}, \orgname{Universidade Federal do Paraná (UFPR)}, \orgaddress{\street{Rua Pioneiro}, \city{Palotina}, \postcode{85950-000}, \state{PR}, \country{Brazil}}}

\affil[7]{\orgdiv{Programa de P\'os-graduação em F\'isica Aplicada}, \orgname{Universidade Federal da Integração Latino-Americana}, \orgaddress{\street{Av. Tarquínio Joslin dos Santos, 1000}, \city{Foz do Iguaçu}, \postcode{85867-670}, \state{PR}, \country{Brazil}}}

\affil[8]{\orgdiv{Departamento de F\'isica, Universidade Federal do Esp\'irito Santo}, \orgname{N\'ucleo de Astrof\'{\i}sica e Cosmologia (Cosmo-Ufes)}, \orgaddress{\street{Av. Fernando Ferrari, 514}, \city{Vit\'oria}, \postcode{29075-910}, \state{ES}, \country{Brazil}}}


\abstract{This study explores the origins of cosmic rays and their secondary messengers, focusing on the potential role of four BL Lacs-W Comae, 1ES 1959+650, PKS 2005-489 and PKS 2155-304-as potential sources of astrophysical neutrinos and gamma rays. We analyzed a single-zone model to understand the interactions between high-energy protons and ambient photons within blazar jets, leading to neutrino production observables and gamma-ray emission. This modeling contextualizes the emissions within multiwavelength observations and evaluates the capabilities of the next-generation Cherenkov Telescope Array Observatory (CTAO) in detecting these emissions. Our estimations suggest that these sources could be effective emitters of CRs, highlighting the need for future multimessenger observations to further investigate and constrain this class of sources.}

\maketitle
\section{Introduction}\label{sec:intro}

Cosmic rays (CRs) that reach Earth with energies greater than $10^{19}$ eV possess a Larmor radius that exceeds the dimensions of the Milky Way and are believed to originate from extreme conditions outside of our Galaxy. The arrival directions determined by the Pierre Auger Observatory indicate a potential correlation between the highest-energy ultra-high-energy cosmic rays (UHECRs, E $>\ 10^{18}$ eV) and nearby active galactic nuclei (AGN) \cite{2018ApJ...853L..29A}, this correlation has not yet been confirmed, and further data are required to establish its certainty. In 1949, the Italian physicist Enrico Fermi (1901-1954) proposed the acceleration mechanism to describe particles accelerated in stochastic collisions \cite{PhysRev.75.1169}; this mechanism could model acceleration in shock waves associated with the remnant of a gravitational collapse, for example, a stellar collapse and the surrounding of a black hole accreting in the center of galaxies \cite{2024PhRvL.132i1401P}, as is the case in blazars \cite{2012ApJ...749...63M}. 

The maximum particle energy achievable during CR acceleration in some TeV blazars can reach $10^{3}$ EeV \cite{2014IJAA....4..499Z}. This energy limit depends on the particle charge $Z$ within an emission region located inside jets, known as ``blob'' and characterized by a radius $R$ and a magnetic field $B$ \cite{1984ARA&A..22..425H,2012ApJ...749...63M, 2014IJAA....4..499Z}. The maximum theoretical acceleration energy $E_\mathrm{max}$, without taking into account efficiency factors, can be expressed as ${E_{\mathrm{max}}}/{10^{15}\mathrm{eV}}$ $\leq \Gamma Z$ ($B$/$\mathrm{\mu}$G)($R$/pc), where $\Gamma$ represents the bulk Lorentz factor of the emitting region. Evidence suggests that jets, powered by supermassive black holes at the center of active galaxies, can accelerate CRs \cite{Zhang2014, Das2022}. The acceleration of CRs is supposed to occur inside a relativistic plasma blob with $\Gamma \sim$ 31.6, suggesting that TeV blazar jets are potential sources for the origin of UHECRs \cite{2014IJAA....4..499Z,2009NJPh...11f5014S}.

UHECRs are deflected from their origin due to the presence of galactic and extragalactic magnetic fields, making it extremely difficult to directly identify their sources \cite{2023arXiv231112120U,2023EPJWC.28304001G}. In addition, the propagation of UHECR also results in the generation of secondary messengers, namely GZK (Greisen-Zatsepin-Kuzmin) gamma rays and neutrinos. The modeling of these interaction processes establishes a connection between particle physics and multimessenger astronomy, enabling the indirect identification of the sources of charged particles. The multimessenger approach has been used to identify high-energy neutrinos as signs of hadronic accelerators \cite{2022ApJ...934..164A, 2024Univ...10..326S}.

Blazars have been discovered in multiwavelength bands, in special radio, X-ray, and gamma rays, and exhibit variability in flux and spectral shape on time scales ranging from years to a few minutes \cite{2021Univ....7..492G,2014IJAA....4..499Z, MAGIC2024}. These objects have long been thought to be likely sources of astrophysical neutrinos \cite{2021Univ....7..492G, 2023ecnp.book..483M,1992A&A...260L...1M}. Statistical searches for neutrino emitters have only recently become feasible due to the accumulation of sufficient data, primarily from the IceCube Neutrino Observatory \cite{Aartsen_2017}. The detection of high-energy astrophysical neutrinos by IceCube has significantly advanced, leading to revolutionary discoveries in particle physics and astrophysics \cite{2018Sci...361.1378I, 2018Sci...361..147I}.
Neutrinos are produced astrophysically through the decay of charged pions by interactions with energetic protons in dense matter or via photoproduction from CR protons that interact with ambient photons. However, a significant portion of the protons may escape the jet if the rate of proton-photon (p$\gamma$) interactions in the jet is less than the escape rate. These CRs protons \textbf{with energy $\gg 1$ EeV} can interact with photons from the extragalactic background light (EBL) and cosmic microwave background (CMB) to create photopions \cite{2020ApJ...889..149D, 2022A&A...658L...6D}. 

The blazars analyzed in this study are drawn from the Candidate Gamma-ray Blazar Survey (CGRaBS) catalog \cite{Healey2008}, which is a flux-limited sample (8.4 GHz flux density $>$ 65 mJy) of 1625 radio-loud AGN identified as potential gamma-ray emitting quasars detectable with Fermi-LAT \cite{2024A&A...681A.119R}. The multiwavelength spectral energy distribution (SED) data for these sources include near-infrared observations from ESO's GROND \cite{Greiner2008}, optical-ultraviolet data from the \textit{Swift} Ultra-Violet Optical Telescope (UVOT) \cite{Roming2005}, and X-ray data from \textit{NuSTAR} \cite{Harrison2013}, \textit{XMM-Newton} \cite{Matthews2001}, \textit{Chandra} \cite{Weisskopf2000}, and the \textit{Swift} X-ray Telescope (XRT) \cite{Burrows2005} and Burst Alert Telescope (BAT) \cite{Krimm2013}, as well as gamma-ray data from the Fermi-LAT \cite{Atwood2009}. For more details on multiwavelength data, see Paliya et al. \cite{2017ApJ...851...33P}.

In this scenario, we reproduced and examined a single-zone model of four BL Lacertae Objects (BL Lacs) - W Comae \cite{2008ApJ...684L..73A, 2009ApJ...707..612A}, 1ES 1959+650 \cite{ 2004ApJ...601..151K, 2005APh....23..537H, 2008ApJ...679.1029T}, PKS 2005-489 \cite{2010MNRAS.401.1570T, 2010A&A...511A..52H} and PKS 2155-304 \cite{2009A&A...502..749A, 2016NewA...44...21B, 2024A&A...681A.119R}, in which emissions originate from electrons and protons within a blob in a jetted configuration \citep{2024A&A...681A.119R}. Specifically, high-energy emission arises from inverse Compton scattering interactions with synchrotron photons generated by the electrons within the blob in blazar jets, while neutrinos are produced via proton-proton (pp) and p$\gamma$ interactions. The paper compares the emission from each source with multiwavelength observations derived from publicly compiled data by Paliya et al. (2017) \citep{2017ApJ...851...33P} and presents the model parameters at the sources. Although direct neutrino emissions from these sources have yet to be observed, they may reveal significant multimessenger emissions in the future through
the analysis of high-energy gamma-ray emissions. This paper demonstrates that the upcoming capabilities of the Cherenkov Telescope Array Observatory (CTAO) offer promising opportunities to detect high-energy gamma-ray emissions in these regions \cite{CTACollaboration}. Such detections would suggest that these blazars can accelerate protons to very high energies, positioning them as promising candidates for investigating potential neutrino production through proton interactions \cite{2013A&A...555A..70T, 2015A&A...578A..32Z, 2018ApJ...866..109S, 2023Galax..11..117A, 2023JHEAp..38....1A}. The detection of gamma rays at high energies is crucial to distinguish between leptonic and hadronic processes. Gamma-ray spectra provide direct evidence that one model over another is favored by revealing distinct spectral characteristics \cite{2020Galax...8...72C, 2022MNRAS.516.1539O}. Furthermore, hadronic origins may show different morphologies \cite{2012ApJ...753...41L, 2023A&A...671A..12M}, and the variability in emissions can differentially affect leptonic and hadronic processes, affecting magnetic fields or particle acceleration mechanisms \cite{2022Galax..10..105S, 2024ApJ...967...93Z}.

Imaging Atmospheric Cherenkov Telescopes (IACTs), which can detect very high-energy gamma rays up to several TeV, have greatly increased what we know about where CRs come from \cite{Postnikov2017, Wagner2022, ACERO2023}. These telescopes can point directly to the gamma-ray sources because gamma rays travel in a straight line from their source without changing direction \cite{2013APh....43...56B, MAGIC, Veritas, HESS}. The CTAO is set to be the main ground-based gamma-ray observatory for the coming decade \cite{2019EPJWC.20901038C}. Additionally, with its distinctive capabilities, the Large High Altitude Air Shower Observatory (LHAASO) \cite{2016NPPP..279..166D, 2023PhRvL.131o1001C} and the Southern Wide-Field Gamma Ray Observatory (SWGO) \cite{2024JInst..19C2065C, 2024MNRAS.531.5061H} will detect the most energetic photons ever recorded and analyze the Galaxy's highest-energy CRs up to the PeV scale \cite{CTACollaboration}. Identifying the extreme accelerators that energize these particles is a CTAO research goal. In this paper, we analyze the regions surrounding four BL Lacs to evaluate the gamma-ray emission detection capabilities of CTAO.

The structure of this paper is as follows. The details of the lepto-hadronic modeling examined are presented in Section~\ref{sec:lp}. In Section~\ref{sec:CTAO}, we provide an in-depth analysis that enables the evaluation of the observability of the blazars by CTAO. The findings of these analyzes indicate a potential correlation between the regions of interest and the emission of lepto-hadronic processes, as described in Section \ref{sec:lp}. Finally, brief remarks close the paper in Section~\ref{sec:conclusions}.

\section{Lepto-hadronic modeling}\label{sec:lp}

This section explores the lepto-hadronic one-zone radiation modeling approach to explain the spectral energy distributions (SEDs) from gamma-ray blazar surroundings. The model assumes that non-thermal relativistic particle populations exist in a spherical emission region in the relativistic jet. This region is defined by its radius, $R$; magnetic field strength, $B$; and the bulk Lorentz factor, $\Gamma$. Particle populations are distributed as a simple power law $dN/d \gamma \propto \gamma^{-\alpha}$, where $N$ is the number of particles, $\alpha$ is the spectral index, and $\Gamma$ is the Lorentz factor. The jet is assumed to be observed at an angle $\theta_{obs} = 1/ \Gamma$ relative to the jet axis, resulting in a Doppler factor $\delta_{D} \approx \Gamma$. The injection luminosity L gives the total power injected into the particles in the radiative zone, which corresponds to the power injected into the CRs, before radiative losses, in the jet rest frame \cite{2024A&A...681A.119R}. This model, which is well described in Rodrigues et al.~\cite{2024A&A...681A.119R}, includes contributions from both leptonic (electron driven) and hadronic (proton driven) processes and provides a comprehensive framework for describing the gamma-ray emissions found in blazars. The radiation mechanisms include the primary electron synchrotron self-Compton (SSC) processes which describe the emission from non-thermal electrons. Synchrotron radiation is generated as these electrons spiral in a magnetic field, while inverse Compton scattering boosts ambient photons to higher energies through interactions with relativistic electrons. The proton SSC processes are key mechanisms for understanding high-energy photon production in astrophysical sources. This process involves relativistic protons radiating synchrotron photons or scattering photons to higher energies through inverse Compton interactions. In addition, particle interactions contribute to high-energy emissions. These include p$\gamma$ interactions—photon-pair production \( {\rm p}\gamma \rightarrow {\rm p} e^- e^+ \) and neutral pion decay \( {\rm p}\gamma \rightarrow \pi^{0} \rightarrow \gamma \gamma \)—and photon-photon ($\gamma\gamma$) pair annihilation \( \gamma\gamma \rightarrow e^- e^+ \). All-flavor p$\gamma$ interactions produce neutrinos of all flavors and other secondary particles.  
  
In the model proposed by Rodrigues et al. (2024) \cite{2024A&A...681A.119R}, X-ray and radio fluxes were treated as upper limits for electron emission. However, in some sources, this emission may reasonably fit the observed data. This approach is based on the fact that the compact radiation zone responsible for high-energy photon and neutrino emission is optically thick at radio frequencies due to synchrotron self-absorption  \cite{2024A&A...681A.119R}. Building on the hadronic interaction framework described in \cite{2024A&A...681A.119R}, we have extended the analysis to include pp interactions. In this process, a highly relativistic proton collides with a low-energy target proton, resulting in the production of neutral and charged pions. These pions decay into high-energy gamma rays and neutrinos - $\pi^0 \rightarrow \gamma\gamma$, $\pi^+ \rightarrow \mu^+ + \nu_\mu$, and $\mu^+ \rightarrow e^+ + \nu_e + \bar{\nu}_\mu$. By incorporating pp interactions, the model captures additional pathways for high-energy gamma-ray and neutrino production, providing a more comprehensive understanding of the physical processes within the jet environments of blazars. This addition provides a more comprehensive modeling approach, capturing contributions from hadronic interactions not previously considered in the original model. All-flavor interactions, including p$\gamma$ and pp processes, contribute significantly to high-energy particle production.

The calculations of the lepto-hadronic model were performed using the open-source AM$^{3}$\footnote{\url{https://am3.readthedocs.io/en/latest/}} software  \cite{Klinger_2024}, which efficiently solves the time-dependent partial differential equations for the energy spectra of electrons, positrons, protons, neutrons, photons, neutrinos, and charged secondaries (pions and muons), immersed in an isotropic magnetic field \cite{AM3}. The model underlying charged and neutral pions from pp interactions is performed with the QGSJET-III Monte Carlo generator of high-energy hadronic collisions \cite{2024PhRvD.109c4002O}. Four BL lacs were analyzed: W Comae, 1ES 1959 + 650, PKS 2005-489, and PKS 2155-304. Table \ref{tab:leptohadronic_modeling} displays the parameters of the lepto-hadronic one-zone
radiation model used. The spectral indices for the proton and electron populations were set equal, with $\alpha_{\rm p} = \alpha_{\rm e} = 1.0$. A constraint was imposed on the maximum proton luminosity, ensuring L$_{\mathrm{p}}/$L$_{\mathrm{e}} <$ 10$^{3}$, along with the corresponding maximum neutrino flux level \cite{2024A&A...681A.119R}.

\begin{table}
\centering
\begin{tabular}{ccccc}
\toprule
 \textbf{Region} & \textbf{W Comae} & \textbf{1ES 1959+650} & \textbf{PKS 2005-489} & \textbf{PKS 2155-304} \\
\midrule
 Blazar class & IBL & HBL & HBL & HBL \\
$\log_{10} R$ (cm) & 16.7 & 16.8 & 16.7 & 17.3 \\
 $\log_{10} \mathrm{B}$ (G)  & 0.7 & 1.4 & 0.7 & 0.9 \\
$\Gamma$ & 5.5 & 8.2 & 12.0 & 4.0 \\
$\log_{10} \mathrm{L^{\rm max}_e}$ \\($\rm{erg\ s^{-1}}$) & 46.2 & 42.0 & 41.4 & 43.5 \\
$\log_{10} \mathrm{L^{max}_{p}}$ \\($\rm{erg\ s^{-1}}$) & $<$46.4 & $<$46.7 & $<$44.1 & $<$47.2 \\
$\log_{10} \mathrm{\Phi^{max}_{\nu}}$ & -13.5 & -11.3 & -14.4 & -11.8 \\
(erg cm$^{-2} s^{-1}$)\\
$z$ & 0.102 & 0.047 & 0.071 & 0.116 \\
\bottomrule
\end{tabular}
\caption{Lepto-hadronic one-zone radiation model parameters (from Rodrigues et al. (2024) \cite{2024A&A...681A.119R}), including blazar class (IBL for intermediate and HBL for high-energy peaked BL Lac objects), emission region radius $R$, magnetic field strength $B$, bulk Lorentz factor $\Gamma$, maximum electron luminosity $\mathrm{L^{max}_e}$, maximum proton luminosity $\mathrm{L^{max}_p}$, maximum neutrino flux $\mathrm{\Phi^{max}_{\nu}}$, and redshift $z$ for the four blazar regions—W Comae, 1ES 1959+650, PKS 2005-489, and PKS 2155-304.}
\label{tab:leptohadronic_modeling}
\end{table}

\begin{figure}[H]
\centering
\includegraphics[width=13.0cm]{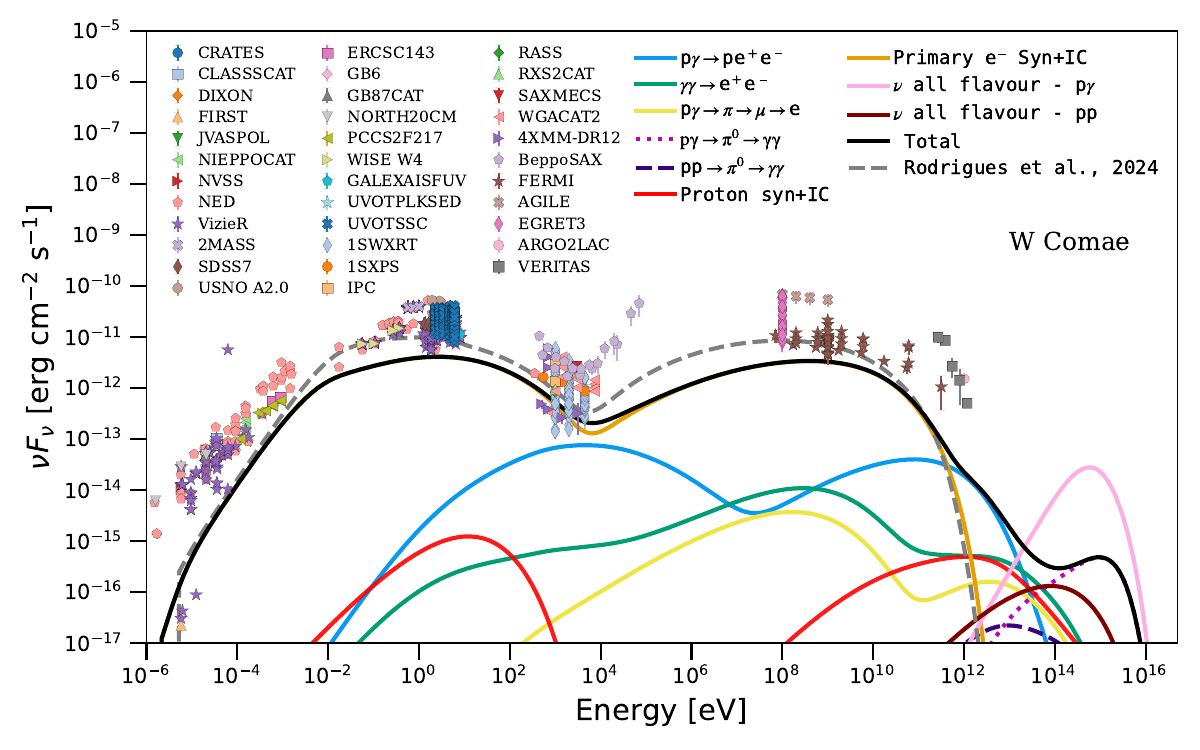}
\caption{Spectral energy distribution (SED) and multiwavelength data of W Comae. The black curve represents the total cumulative SED from all processes in the lepto-hadronic model. The black curve depicts the total cumulative SED from all processes in the lepto-hadronic model, representing the fit achieved in this work. The gray curve shows the model from Rodrigues et al. (2024) \cite{2024A&A...681A.119R}. In comparison to the purely leptonic model, the lepto-hadronic model is characterized by the dominant contributions to the emissions of high-energy photons and neutrinos. These processes originate from the following interactions: proton-photon interactions (p$\gamma$, dotted pink curve) are observed from $10^{13}$ eV to $10^{16}$ eV. Pion production from proton-proton interactions (pp → $\pi^{0}$, dark blue dashed curve) is mainly observed from approximately $10^{12}$ to $10^{14}$ eV, while proton-driven SSC processes (red curve) are significant from $10^{8}$ to $10^{14}$ eV. The all-flavor neutrino fluxes (pink and brown curves) exhibit a peak near $10^{14}$ eV and extend to $10^{16}$ eV (pink) and $10^{15}$ eV (brown).}
\label{fig:wcomae}
\end{figure}

\begin{figure}[H]
\centering
\includegraphics[width=13.0cm]{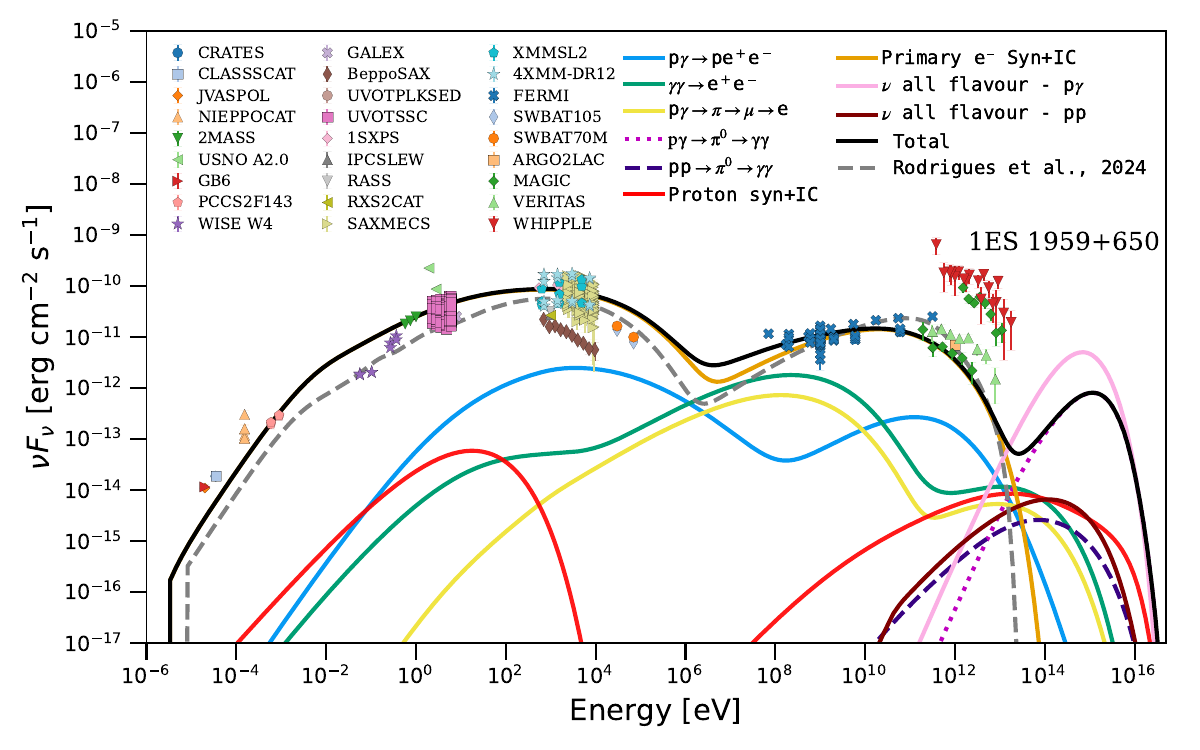}
\caption{Spectral energy distribution (SED) and multiwavelength data of 1ES1959+650. The black curve represents the total cumulative SED from all processes in the lepto-hadronic model. The black curve depicts the total cumulative SED from all processes in the lepto-hadronic model, representing the fit achieved in this work. The gray curve shows the model from Rodrigues et al. (2024) \cite{2024A&A...681A.119R}. The processes from the lepto-hadronic model are: proton-photon interactions (p$\gamma$, dotted pink curve) are observed from $10^{12}$ eV to $10^{16}$ eV. Pion production from proton-proton interactions (pp → $\pi^{0}$, dark blue dashed curve) is mainly seen from approximately $10^{10}$ to $10^{16}$ eV, while proton-driven SSC processes (red curve) are significant from $10^{8}$ to $10^{16}$ eV. The all-flavor neutrino fluxes (pink and brown curves) exhibit a peak near $10^{14}$ eV and extend to $10^{16}$ eV (pink and brown).}
\label{fig:1ES1959}
\end{figure}

\begin{figure}[H]
\centering
\includegraphics[width=13.0cm]{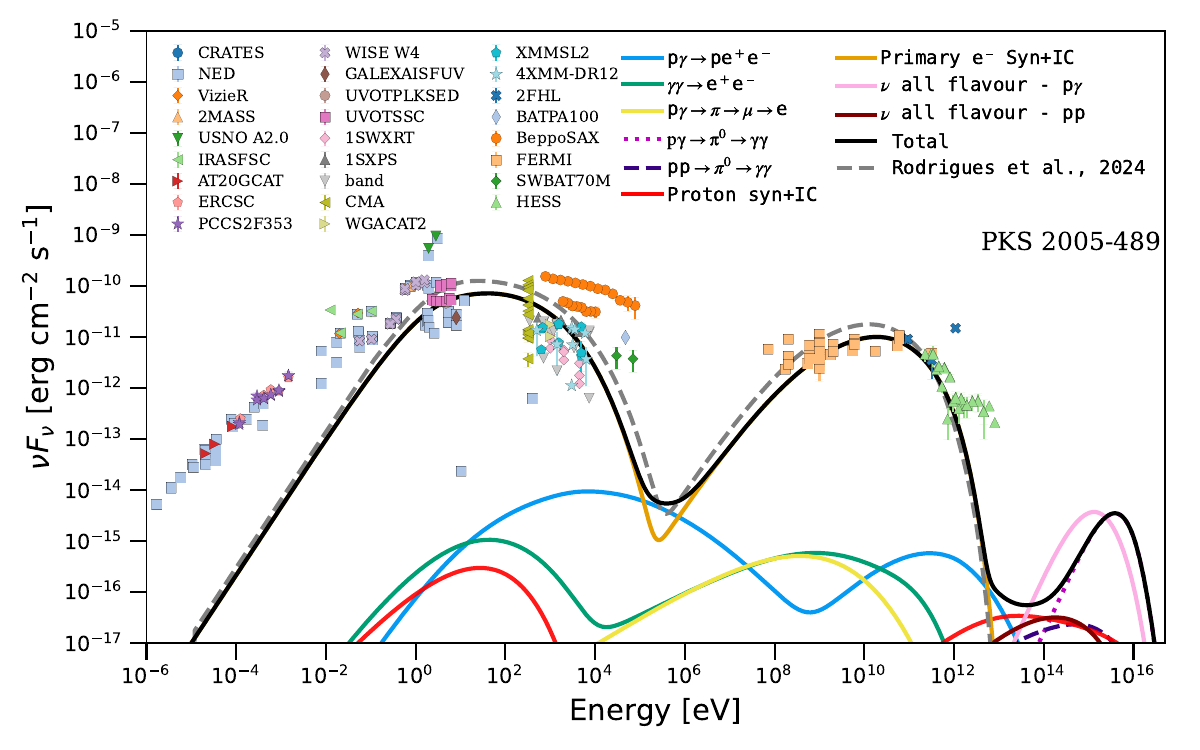}
\caption{Spectral energy distribution (SED) and multiwavelength data of PKS 2005-489. The black curve represents the total cumulative SED from all processes in the lepto-hadronic model. The black curve depicts the total cumulative SED from all processes in the lepto-hadronic model, representing the fit achieved in this work. The gray curve shows the model from Rodrigues et al. (2024) \cite{2024A&A...681A.119R}. The lepto-hadronic model is characterized by the dominant contributions to the emissions of high-energy photons and neutrinos. The lepto-hadronic processes result from the following interactions: proton-photon interactions (p$\gamma$, dotted pink curve) are observed from $10^{14}$ eV to $10^{16}$ eV. Pion production from proton-proton interactions (pp → $\pi^{0}$, dark blue dashed curve) is mainly seen from approximately $10^{13}$ to $10^{15}$ eV, while proton-driven SSC processes (red curve) are significant from $10^{12}$ to $10^{16}$ eV. The all-flavor neutrino fluxes (pink and brown curves) exhibit a peak near $10^{15}$ eV (pink) and $10^{14}$ eV (brown), extended to $10^{16}$ eV (pink) and $10^{15}$ eV (brown).
}
\label{fig:PKS2005}
\end{figure}

\begin{figure}[H]
\centering
\includegraphics[width=13.0cm]{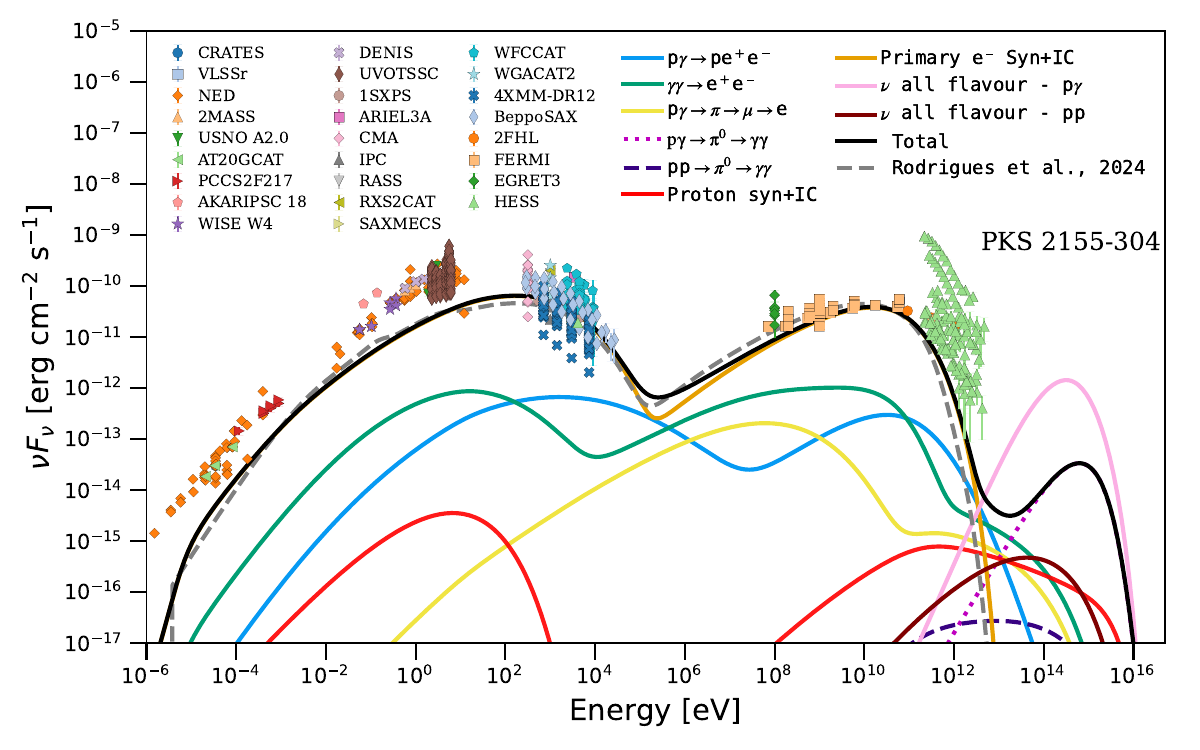}
\caption{Spectral energy distribution (SED) and multiwavelength data of PKS 2155-304. The black curve represents the total cumulative SED from all processes in the lepto-hadronic model. The black curve depicts the total cumulative SED from all processes in the lepto-hadronic model, representing the fit achieved in this work. The gray curve shows the model from Rodrigues et al. (2024) \cite{2024A&A...681A.119R}. Lepto-hadronic processes arise from the following interactions: proton-photon interactions (p$\gamma$, dotted pink curve) are observed from $10^{12}$ eV to $10^{16}$ eV. Pion production from proton-proton interactions (pp → $\pi^{0}$, dark blue dashed curve) is mainly seen from approximately $10^{11}$ to $10^{14}$ eV, while proton-driven SSC processes (red curve) are significant from $10^{8}$ to almost $10^{16}$ eV. The all-flavor neutrino fluxes (pink and brown curves) exhibit a peak near $10^{14}$ eV and extend to $10^{16}$ eV (pink) and $10^{15}$ eV (brown).}
\label{fig:PKS2155}
\end{figure}

Figure \ref{fig:wcomae} displays the multiwavelength SED for the blazar W Comae along with observational data from several catalogs. The thick black line denotes the total fit of the model. The model suggests a possible hadronic contribution based on high-energy data from the VERITAS experiment \cite{Acciari_2008, Biteau_2015} (E $\sim$ $10^{12}$ eV); at this energy, purely leptonic emission from accelerated electrons (orange curve) does not account for the flux. The SED for the blazar 1ES 1959 + 650 is shown in Figure \ref{fig:1ES1959}. The plot shows a better fit to the data, indicating a greater contribution of gamma rays from hadronic processes (E $\sim$ $10^{10}$ - $10^{14}$ eV). At high-energy, both quiescent and flare components are visible. Although the flare data do not align with the model fit, they are included to illustrate the activity of the blazar. The hadronic model for the blazar PKS 2005-489, in figure \ref{fig:PKS2005}, has no significant impact, indicating that the data are well explained solely by emission from accelerated electrons (synchrotron + inverse Compton scattering - orange curve). However, HESS measured fluxes in the range of E $\sim$ $10^{11}$ - $10^{13}$ eV \cite{ HessPKS2005,Biteau_2015}, indicating a potential contribution from the lepto-hadronic model at high energies. This discussion will be explored further in the next section with projections for the CTAO \cite{CTACollaboration}. Finally, the figure \ref{fig:PKS2155} shows the SED of the blazar PKS 2155-304. This blazar presents an intense variability that leads to a significant and widespread dispersion of its multiwavelength fluxes, already discussed in \cite{2024A&A...681A.119R}. Similarly to PKS 2005-489, the hadronic contribution is small but present. High-energy flares require a lepto-hadronic model that incorporates contributions from various processes.
The analysis of these sources indicates the need for a lepto-hadronic approach at high energies, which may suggest future neutrino detections and, consequently, evidence of nucleus acceleration in these sources \cite{2018ApJ...866..109S, 2023Galax..11..117A, 2023JHEAp..38....1A}. The next section will describe the observability of the blazars by the CTAO. Future investigations and more sensitive detections such as CTAO \cite{CTACollaboration} will be essential to confirm these sources and potentially discover new ones, expanding our understanding of high-energy cosmic phenomena \cite{2023ApJ...954...75A,CTACollaboration}.


\section{Sensitivity of CTAO to gamma-ray emission}\label{sec:CTAO}

In this section, we use two approaches from the study by \cite{Costa2024} to evaluate the CTAO's ability to detect gamma-ray emissions from the blazar surroundings. First, Section~\ref{sec:CTAO.1}, we perform a simultaneous likelihood fit using data from multiple catalogs to construct a source model covering the multi-GeV to the multi-TeV range. Then, Section~\ref{sec:CTAO.2}, we apply the 1D ON/OFF observation technique to assess the anticipated performance of CTAO when observing the modeled source \cite{Donath_2023}. We used the Gammapy software\footnote{\url{https://gammapy.org/}} \citep{Deil2017} to perform these analyses.  


\subsection{Source modeling}\label{sec:CTAO.1}

We used data from previous observations of electromagnetic counterparts at a radius $< 0.11^{\circ}$ from the source position to determine the source model in the multi-GeV to multi-TeV range. Specifically, we utilized the VERITAS Catalog of Gamma-Ray
Observations (VTSCat) \cite{Acharyya_2023}, the 3FGL \cite{Acero_2015} and 4FGL \cite{Abdollahi2022} catalogs, corresponding to the $3^{\rm{rd}}$ and $4^{\rm{th}}$ Fermi gamma-ray LAT source catalogs, respectively, and the 2FHL \cite{Ackermann_2016} and 3FHL \cite{Ajello2017} catalogs, representing the $2^{\rm{nd}}$ and $3^{\rm{rd}}$ hard Fermi-LAT source catalogs. Four blazar surroundings were examined in the modeling:
\begin{itemize}
    \item \textbf{W Comae region}: The W Comae region is associated with the well-known BL Lac W Comae. The 3FGL \cite{Acero_2015} and 4FGL \cite{ballet2023fermi, Abdollahi2022} catalogs provide data in the low-energy gamma-ray range, while higher-energy data from the 3FHL \cite{Ajello2017} catalog offers additional spatial and spectral detail. Table~\ref{tab:WComae} displays the celestial coordinates and angular separations from the region's center of the counterparts utilized in the simultaneous likelihood fit, demonstrating precise position alignment across catalog sources.
\end{itemize}
    \begin{table}[h!]
    \centering
    \begin{tabular}{llll}
    \toprule
    \textbf{Counterparts} & \textbf{R.A. (deg)} & \textbf{Dec. (deg)} & \textbf{Sep. (deg)} \\
    \midrule
    3FGL J1221.4+2814 \cite{Acero_2015} & 185.37 & 28.24 & 0.010 \\
    4FGL J1221.5+2814 \cite{ballet2023fermi, Abdollahi2022} & 185.38 & 28.22 & 0.006 \\
    3FHL J1221.5+2813 \cite{Ajello2017} & 185.38 & 28.23 & 0.003 \\

    \bottomrule
    \end{tabular} 
    \caption{Celestial coordinates in the International Celestial Reference System (ICRS) using right ascension (R.A.) and declination (Dec.), along with the angular separation (Sep.) of the counterparts for the  W Comae region centered at (185.38$^\circ$, 28.23$^\circ$) \cite{Truebenbach_2017}.}
    \label{tab:WComae}
    \end{table}
\begin{itemize}
    \item \textbf{1ES 1959+650 region}: This region is associated with the BL Lac 1ES 1959+650, a well-studied blazar known for emissions across multiple wavelengths, including observations from VTSCat \cite{Acharyya_2023}, which has provided valuable very-high-energy (VHE) gamma-ray data for this source (see Figure~\ref{fig:1ES1959}). The 3FGL \cite{Acero_2015} and 4FGL \cite{ballet2023fermi, Abdollahi2022} catalogs capture low-energy gamma-ray data, while the 2FHL \cite{Ackermann_2016} and 3FHL \cite{Ajello2017} catalogs extend coverage to higher energies. As shown in Table~\ref{tab:1ES}, the low angular separations among counterparts used in the simultaneous likelihood fit indicate a stable source position across catalogs, supporting a comprehensive analysis of the spectral features and variability of 1ES 1959+650, including insights into its behavior at VHE through VERITAS observations.
\end{itemize}
    \begin{table}[h!]
    \centering
    \begin{tabular}{llll}
    \toprule
    \textbf{Counterparts} & \textbf{R.A. (deg)} & \textbf{Dec. (deg)} & \textbf{Sep. (deg)} \\
    \midrule
    3FGL J2000.0+6509 \cite{Acero_2015} & 300.02 & 65.15 & 0.009 \\
    4FGL J2000.0+6508 \cite{ballet2023fermi, Abdollahi2022} & 300.01 & 65.15 & 0.005 \\
    2FHL J2000.1+6508 \cite{Ackermann_2016} & 300.03 & 65.14 & 0.014 \\
    3FHL J1959.9+6508 \cite{Ajello2017} & 300.00 & 65.15 & 0.001 \\
    VER J1959+651 \cite{Acharyya_2023} & 300.00 & 65.15 & 0.00\\
    \bottomrule
    \end{tabular} 
    \caption{Celestial coordinates in the International Celestial Reference System (ICRS) using right ascension (R.A.) and declination (Dec.), along with the angular separation (Sep.) of the counterparts for the  1ES 1959+650 region centered at (300.00$^\circ$, 65.15$^\circ$) \cite{Hunt_2021}.}
    \label{tab:1ES}
    \end{table}
\newpage
\begin{itemize}
    \item \textbf{PKS 2005-489 region}: This region corresponds to the BL Lac PKS 2005-489, which emits high-energy radiation. The 3FGL \cite{Acero_2015} and 4FGL \cite{ballet2023fermi, Abdollahi2022} catalogs cover the lower-energy range, while the 2FHL \cite{Ackermann_2016} and 3FHL \cite{Ajello2017} catalogs add high-energy data. Table~\ref{tab:pks2005} shows consistent angular separations of the counterparts across catalogs, aiding precise modeling of PKS 2005-489’s spectral features.
\end{itemize}
    \begin{table}[h!]
    \centering
    \begin{tabular}{llll}
    \toprule
    \textbf{Counterparts} & \textbf{R.A. (deg)} & \textbf{Dec. (deg)} & \textbf{Sep. (deg)} \\
    \midrule
    3FGL J2009.3-4849 \cite{Acero_2015} & 302.35 & -48.83 & 0.006 \\
    4FGL J2009.4-4849 \cite{ballet2023fermi, Abdollahi2022} & 302.36 & -48.82 & 0.007 \\
    2FHL J2009.4-4849 \cite{Ackermann_2016} & 302.36 & -48.82 & 0.009 \\
    3FHL J2009.4-4849 \cite{Ajello2017} & 302.36 & -48.82 & 0.015 \\

    \bottomrule
    \end{tabular} 
    \caption{Celestial coordinates in the International Celestial Reference System (ICRS) using right ascension (R.A.) and declination (Dec.), along with the angular separation (Sep.) of the counterparts for the  PKS 2005-489 region centered at (302.36$^\circ$, -48.83$^\circ$)~\cite{Xu_2019}.}
    \label{tab:pks2005}
    \end{table}
\begin{itemize}
    \item \textbf{PKS 2155-304 region}: The PKS 2155-304 region is a prominent BL Lac with high-energy emissions. This region is represented across multiple Fermi LAT catalogs, including the 3FGL \cite{Acero_2015} and 4FGL \cite{ballet2023fermi, Abdollahi2022} catalogs for low-energy data, and the 2FHL \cite{Ackermann_2016} and 3FHL \cite{Ajello2017} catalogs for high-energy data. Angular separations among catalog entries are minimal (Table~\ref{tab:pks2155}), highlighting stable positioning and emission characteristics.
\end{itemize}

    \begin{table}[h!]
    \centering
    \begin{tabular}{llll}
    \toprule
    \textbf{Counterparts} & \textbf{R.A. (deg)} & \textbf{Dec. (deg)} & \textbf{Sep. (deg)} \\
    \midrule
    3FGL J2158.8-3013 \cite{Acero_2015}&   329.72 &     -30.23 &       0.003 \\
    4FGL J2158.8-3013  \cite{ballet2023fermi, Abdollahi2022}&   329.71 &     -30.23 &       0.002 \\
    2FHL J2158.8-3013 \cite{Ackermann_2016}&   329.72 &     -30.22 &       0.002 \\
    3FHL J2158.8-3013 \cite{Ajello2017}&   329.72 &     -30.22 &       0.001 \\
    \bottomrule
    \end{tabular} 
    \caption{Celestial coordinates in the International Celestial Reference System (ICRS) using right ascension (R.A.) and declination (Dec.), along with the angular separation (Sep.) of the counterparts for the  PKS 2155-304 region centered at (329.72$^\circ$, -30.23$^\circ$) \cite{LeBail_2016}.}
    \label{tab:pks2155}
    \end{table}
\newpage
SEDs are defined by:
\begin{equation}\label{eq:obs}
\Phi (E)_{\rm{obs}} = \Phi (E)_{\rm{int}}~e^{-\tau (E, z)},     
\end{equation}
where $\Phi (E)_{\rm{obs}}$ and $\Phi (E)_{\rm{int}}$ are the observed and intrinsic source spectra, respectively, and $\tau$ is the optical depth, based on the Saldana-Lopez EBL model \cite{Saldana_Lopez_2021} and dependent on gamma-ray energy, $E$, and source redshift, $z$. 

In order to analyze the intrinsic spectral shape, we tested three models:
\begin{itemize}
    \item \textbf{Power Law (PL)}:
    \begin{equation}\label{eq:PL}
    \Phi(E)  = \Phi_{0} \left(\frac{E}{E_0}\right) ^ {-\alpha},
    \end{equation}
    where $\Phi_{0}$ is the amplitude, $E_0 = 1~\rm{TeV}$ is the reference energy, and $\alpha$ is the index.
    
    \item \textbf{Log Parabola (LP)}:
    \begin{equation}\label{eq:LP}
    \Phi(E) = \Phi_0 \left(\frac{E}{E_{0}}\right)^{- \alpha - \beta \log{\left (\frac{E}{E_{0}} \right )}},
    \end{equation}
    where $\beta$ is the index parameter scale.
    
    \item \textbf{Exponential Cutoff Power Law (ECPL)}:
    \begin{equation}\label{eq:ECPL}
    \Phi(E) = \Phi_0 \left(\frac{E}{E_{0}}\right)^{- \alpha} \exp{\left (- \frac{E}{E_{\rm cut}} \right )},
    \end{equation}
    where $E_{\rm cut}$ is the cutoff energy.
\end{itemize}
We utilized the Akaike Information Criterion (AIC) \cite{Akaike1974} to pick the optimal model for the data. The AIC is a model selection criterion that assesses goodness-of-fit while punishing model complexity to prevent overfitting.
Table \ref{tab:AIC} compares the effectiveness of the LP and BPL spectral models (see Equations~\ref{eq:LP} and \ref{eq:ECPL}, respectively) relative to the PL model (see Equation~\ref{eq:PL}) for four blazar regions. The table displays the values of $\delta \rm{AIC_{LP}}$ and $\delta \rm{AIC_{BPL}}$, which represent the percentage difference in AIC values for each model in comparison to the PL model. This efficiency metric, defined as $\delta \rm{AIC_{m} = (1 - BIC_{m} /AIC_{PL}})100\%$, with (m) representing either LP or BPL, determines how well each model fits the data relative to the PL model. The results show that for all regions, both LP and BPL models provide a better fit than the PL model, with varying degrees of improvement. Notably, PKS 2155-304 exhibits the highest $\delta \rm{AIC_{LP}}$ and $\delta \rm{AIC_{BPL}}$ values, indicating a significant improvement in model efficiency with LP and BPL compared to PL. For PKS 2155-304, the LP model efficiency reaches $87.50\%$, and the BPL model achieves $63.99\%$, suggesting that these more complex models capture the spectral characteristics of this source better than the simpler PL model. The other regions, while showing a lesser degree of improvement, still benefit from the LP and BPL models, W Comae has the lowest $\delta \rm{AIC}$ values, indicating a relatively small improvement over the PL model. Overall, the results in Table \ref{tab:AIC} illustrate that the LP model generally provides a better fit than the PL and BPL models across the regions analyzed, except in certain cases where the BPL model also offers competitive improvements.

\begin{table*}[h!]
\begin{minipage}{\textwidth} 
\centering 
\begin{tabular}{lllll} 
\toprule 
\textbf{Region}	& \textbf{W Comae} & \textbf{1ES 1959+650} & \textbf{PKS 2005-489} & \textbf{PKS 2155-304}\\
\midrule
$\delta\ \rm{AIC_{LP}}\footnote{$\delta\ \rm{AIC_{m} = (1 - AIC_{m} /AIC_{PL}})100\%$, where the subscript (m) indicates LP or BPL.}$
& 17.74\% & 34.24\% & 39.70\% &87.50\% \\
$\delta\ \rm{AIC_{BPL}}$
& 12.70\% & 23.47\% & 29.58\% & 63.99\% \\
\bottomrule
\end{tabular}
\caption{Comparison of spectral model efficiencies for four blazar regions—W Comae ($z = 0.102$), 1ES 1959+650 ($z = 0.047$), PKS 2005-489 ($z = 0.071$), and PKS 2155-304 ($z = 0.117$)—using log-parabola (LP) and exponential cutoff power law (ECPL) models (see Equations~\ref{eq:LP} and \ref{eq:ECPL}, respectively) relative to the power law (PL) model (see Equation~\ref{eq:PL}). The spectral models include absorption effects based on the Saldana-Lopez EBL model \cite{Saldana_Lopez_2021} (see Equation~\ref{eq:obs}). Model performance is evaluated using the Akaike Information Criterion (AIC \cite{Akaike1974}), with $\delta \rm{AIC_{LP}}$ and $\delta \rm{AIC_{BPL}}$ indicating the relative efficiencies of the LP and BPL models compared to the PL model.}
\label{tab:AIC}
\end{minipage}
\end{table*}

Table \ref{tab:sourse_models} presents the spectral model parameters for each blazar region, as derived from the simultaneous likelihood fit. The table includes key parameters of the modeled sources, such as celestial coordinates and redshift. These details define the source models used for the four blazar regions—W Comae, 1ES 1959+650, PKS 2005-489, and PKS 2155–304.

\begingroup
\renewcommand*{\thefootnote}{\alph{footnote}}  
\begin{table}[h!]
\centering
\begin{minipage}{\textwidth}
    \centering
\begin{tabular}{ccccc}
\toprule
\textbf{Region}	& \textbf{W Comae} & \textbf{1ES 1959+650} & \textbf{PKS 2005-489} & \textbf{PKS 2155-304} \\
\midrule

$\Phi_0 \times 10^{-12}$ & $0.563 \pm 0.198$ & $9.40 \pm 0.60$ & $3.26 \pm 0.74$ & $15.1 \pm 1.4$ \\ 
($\mathrm{cm^{-2}\,s^{-1}\,TeV^{-1}}$)  \\

$\alpha$ & 
$2.46 \pm 0.11$ & $2.14 \pm 0.03$ & $2.18 \pm 0.09$ & $2.28 \pm 0.03$ \\

$\beta \times 10^{-2}$ & 
$2.57 \pm 0.90$ & $2.70 \pm 0.26$ & $3.29 \pm 0.79$ & $3.83 \pm 0.28$ \\

Position (R.A., Dec.) & $185.38, 28.23$ & $300.00, 65.15$ & $302.36, -48.83$ & $329.72, -30.22$ \\
(deg)\\

$z$ & 
$0.102$ & $0.047$ & $0.071$ & $0.117$ \\
\bottomrule
\end{tabular}
\caption{Spectral model parameters, amplitude $\Phi_0$ (scaled by $10^{-12}$), spectral index $\alpha$ and curvature parameter $\beta$ (scaled by $10^{-2}$), including 
celestial coordinates in the International Celestial Reference System (ICRS) using right ascension (R.A.) and declination (Dec.) and redshift, $z$, for the four blazar regions—W Comae, 1ES 1959+650, PKS 2005-489, and PKS 2155-304. The gamma-ray spectra were modeled using a log parabola (LP) model (see Equation~\ref{eq:LP}), with parameters derived from the Saldana-Lopez EBL model \cite{Saldana_Lopez_2021}  for extragalactic background light (EBL) absorption (see Equation~\ref{eq:obs}).}
\label{tab:sourse_models}
\end{minipage}
\end{table}
\endgroup

\newpage
\subsection{Expected Observational Performance of CTAO}\label{sec:CTAO.2}

In order to evaluate the CTAO's performance in monitoring the modeled sources (see Table~\ref{tab:sourse_models}), we employed the well-known ON/OFF observational approach in gamma-ray astronomy, specifically designed for Cherenkov telescopes. This method, described in detail in \cite{Piano2021} and referenced in recent studies such as \cite{Costa2024}, allows for emission assessments by distinguishing between ON (source-centered) and OFF (background) regions. This approach enables accurate background estimation, essential for calculating photon excess and detection significance. The ON region was chosen by taking a simple circular region with the center at the source position and a radius of 0.11$^\circ$. We used the Instrument Response Functions (IRFs) provided by the CTA Consortium and CTAO (version prod5 v0.1 \cite{CTAOIRFS}), assuming parallel pointing for all telescopes and setting the telescopes in wobbling mode around the source position with a 0.5$^\circ$ offset.

The IRFs, configured in the ``Alpha Configuration'', cover both CTAO’s North and South arrays and include specific subarray configurations: CTAO North with 4 LSTs (Large-Sized Telescopes) and 9 MSTs (Medium-Sized Telescopes); CTAO North-LSTs with 4 LSTs and CTAO North-MSTs with 9 MSTs and CTAO South with 14 MSTs and 37 SSTs (Small-Sized Telescopes); CTAO South-MSTs with 14 MSTs and CTAO South-SSTs with 37 SSTs. Performance estimates were calculated for three zenith angles (20$^\circ$, 40$^\circ$, and 60$^\circ$), using azimuth-averaged IRFs optimized for a 50-hour observation time.
Background values were determined using IRF templates, with a background scaling factor of $\alpha = 0.2$, aligned with IRF specifications for consistent sensitivity evaluations. We set a minimum of 10 expected signal counts and a minimum significance of $5\ \sigma$ per bin \cite{CTAOIRFS}.

To establish the optimal location of the array (CTAO North or CTAO South) and the zenith angle (20$^\circ$, 40$^\circ$, or 60$^\circ$) for observing the modeled sources, we calculated the annual visibility time for each source of each array.  We considered three zenith angle ranges — 10$^\circ$ – 30$^\circ$, 30$^\circ$ – 50$^\circ$, and 50$^\circ$ – 70$^\circ$ — corresponding to IRFs optimized for zenith angles of 20$^\circ$, 40$^\circ$, and 60$^\circ$, respectively. For a source to be considered visible, its zenith angle must lie between a minimum and maximum altitude, with lower zenith angles indicating that the source is closer to the observer's zenith (directly overhead), providing better observation conditions. The visibility calculations accounted for each source's observability during typical nighttime hours (from 18:00 to 06:00 the following morning) throughout 2025, with a time step of 30 minutes. Based on these visibility results, we generated energy-dependent differential flux sensitivity curves for both full array and subarray IRF configurations, assuming a 50-hour observation time. The integral sensitivity of each curve was then calculated and the configuration that covered the energy range of interest with the lowest value was chosen to simulate CTAO observations. Gamma rays were generated within the energy range of 100 GeV to 32 TeV, which is narrower than the intended energy range for CTAO in full array setups \cite{CTAOIRFS}. In the following, we present the visibility results and corresponding analyses for the four blazar regions.

\begin{itemize}
    \item \textbf{W Comae Region}: As Table~\ref{tab:WComae_visibility} displays, only the 60$^\circ$ zenith angle offers visibility (1163 hours) from the CTAO South for the W Comae region. Although 60$^\circ$ is not ideal due to the greater atmospheric absorption, it remains the only option for observing this region from the South. In contrast, the CTAO North provides visibility at 20$^\circ$, 40$^\circ$, and 60$^\circ$ zenith angles, with the 20$^\circ$ IRF offering 556.5 hours of observation. Despite significantly lesser visibility, the CTAO North (as indicated in \cite{CTACollaboration} - Table 12.1) and 20$^\circ$ zenith angle are preferable for improved observational quality. Figure~\ref{fig:WComae_arrays}  displays energy spectra without (intrinsic) and with EBL attenuation \cite{Saldana_Lopez_2021}. Additionally, it shows the differential sensitivity curves optimized for various CTAO North configurations: the full array with an integral sensitivity of $1.11 \times 10^{-12}\ \rm{cm^{-1}} s^{-1}$, CTAO North-LSTs ($1.54 \times 10^{-12}\ \rm{cm^{-1}} s^{-1}$), and CTAO North-MSTs ($1.38 \times 10^{-12}\ \rm{cm^{-1}} s^{-1}$). Figure \ref{fig:WComae_sed} presents the SED, including CTAO's observations and the flux points of the counterparts used in the likelihood analysis (see Table~\ref{tab:WComae}).

    \begin{table*}[h!]
    \begin{minipage}{\textwidth}
    \centering
    \begin{tabular}{lll}
    \toprule
     \textbf{Array} & \textbf{Zenith} & \textbf{Visibility} \\
    \midrule
    CTAO South& 20$^\circ$ & - \\
    CTAO South & 40$^\circ$ & - \\
    CTAO South & 60$^\circ$ & 1163.00 h\\
    CTAO North & 20$^\circ$ & 556.50 h\\
    CTAO North & 40$^\circ$ & 565.00 h\\
    CTAO North & 60$^\circ$ & 583.50 h\\
    \bottomrule
    \end{tabular}
    \caption{Visibility of the W Comae region from the CTAO South and North arrays at different zenith angles. Calculations are based on 2025 during typical nighttime hours (from 18:00 to 06:00 the following morning) with a time step of 30 minutes. Visibility is given in hours and dashes indicate the source is not visible at the corresponding array and zenith angle.}
    \label{tab:WComae_visibility}
    \end{minipage}
    \end{table*}

    \begin{figure}[h!]
    \centering
    {\includegraphics[angle=0,width=0.8\textwidth]{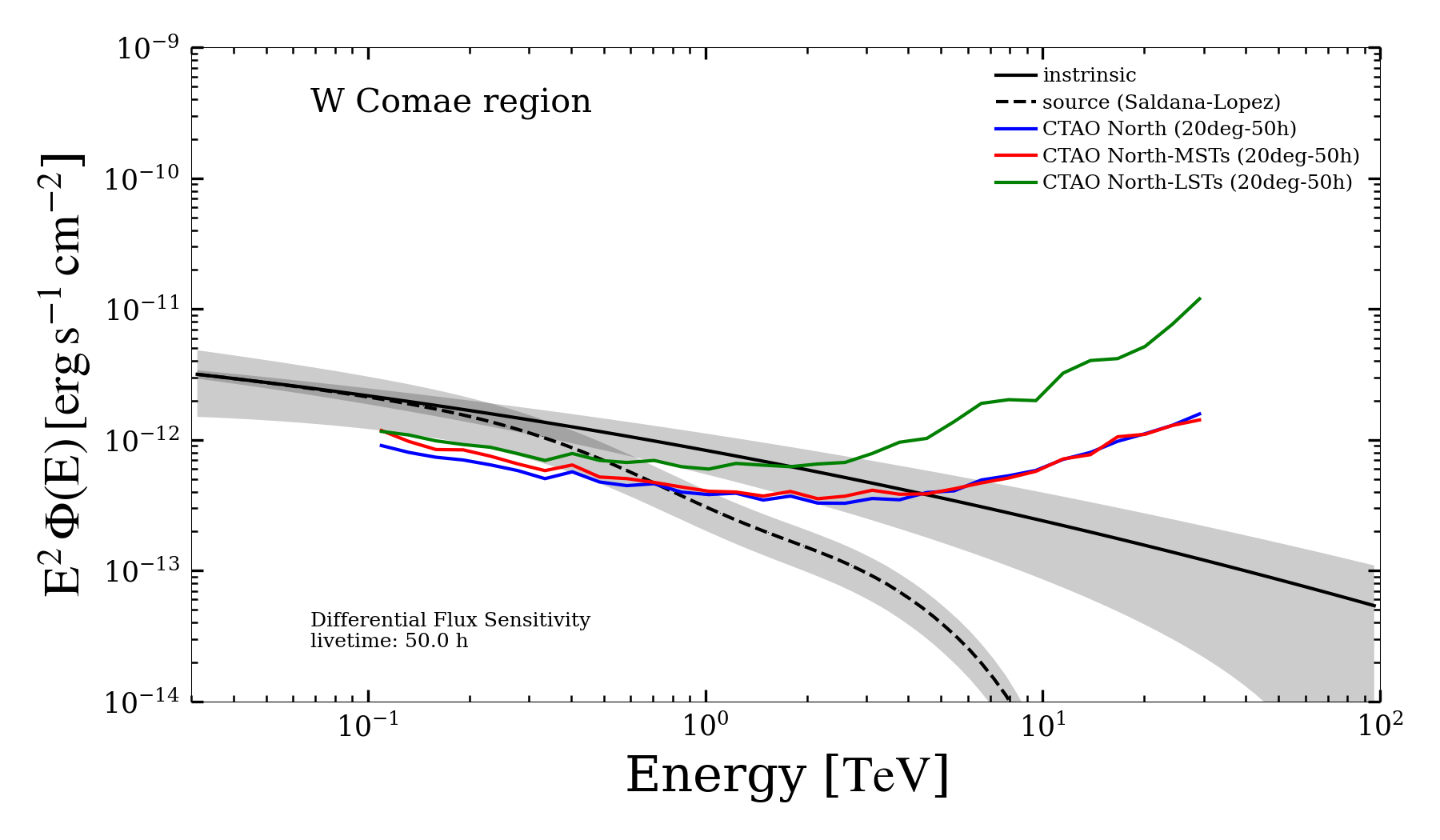}}
    \caption{
        Energy-dependent differential flux sensitivity and spectral model for the W Comae region. The black solid line represents the intrinsic model, while the black dashed line shows the \textbf{absorbed spectrum} model by Saldana-Lopez  \cite{Saldana_Lopez_2021} (see Table~\ref{tab:sourse_models}). The solid-colored lines display sensitivity curves for the CTAO North with different array configurations: (blue line) CTAO North with 4 LSTs (Large-Sized Telescopes) and 9 MSTs (Medium-Sized Telescopes); (red line)  CTAO North-MSTs with 9 MSTs and (green line) CTAO North-LSTs with 4 LSTs. The IRFs are optimized to 20$^{\circ}$ zenith angle and the differential flux sensitivity refers to a 50-hour observation time.}
    \label{fig:WComae_arrays}
    \end{figure}
    
    \begin{figure}[H]
    \centering
    {\includegraphics[angle=0,width=0.8\textwidth]{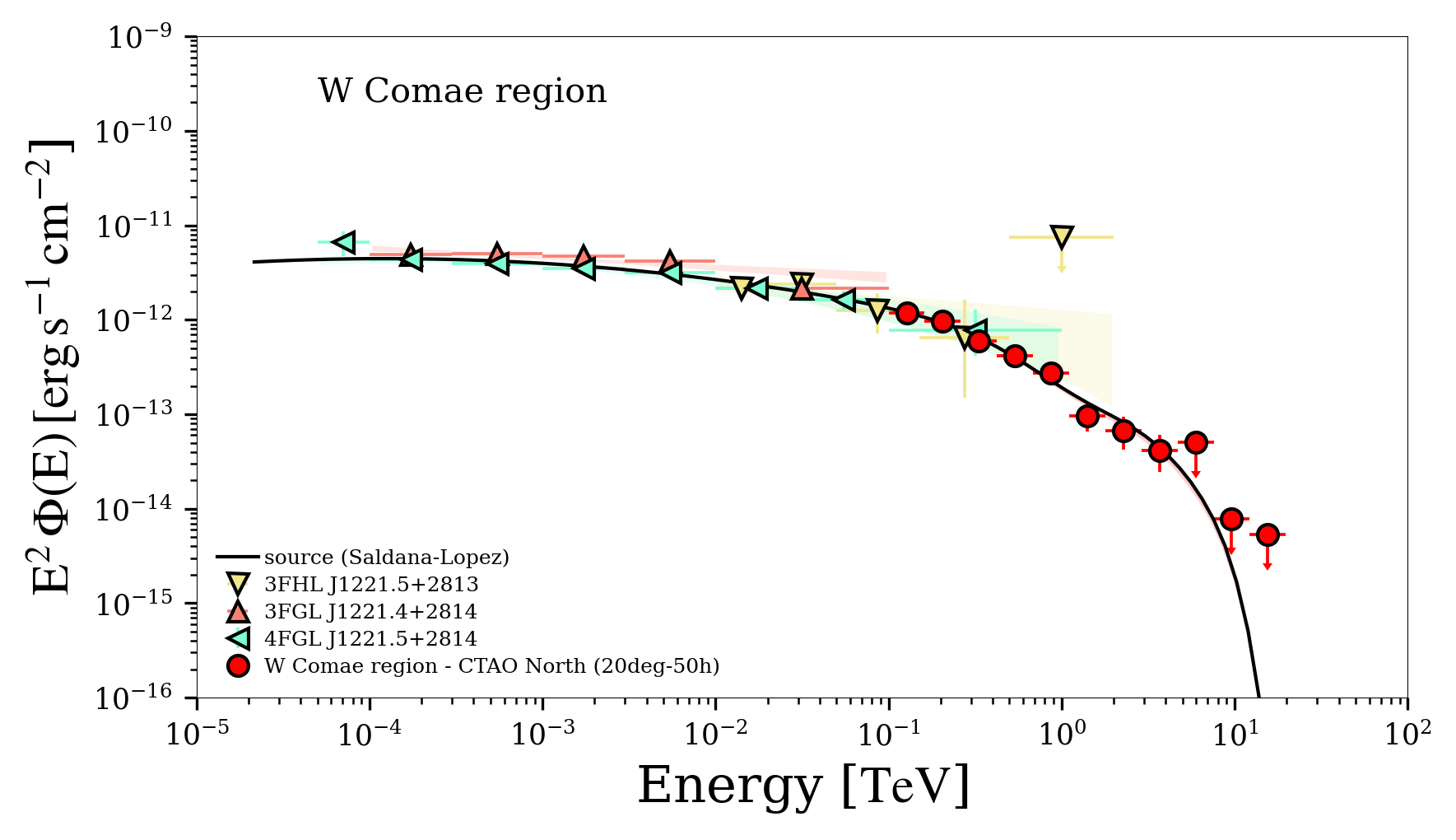}}
    \caption{Spectral energy distribution (SED) for the W Comae region. This plot displays the modeled gamma-ray flux for W Comae, overlaid with observational data from Fermi-LAT catalogs: 3FHL, 3FGL, and 4FGL (see Table~\ref{tab:WComae}). The black solid line represents the absorbed spectrum model by Saldana-Lopez \cite{Saldana_Lopez_2021} (see Table~\ref{tab:sourse_models}) and the red circles show the flux points from the CTAO South array for a 50-hour observation. Error bars indicate the statistical uncertainties for each data point.}
    \label{fig:WComae_sed}
    \end{figure}
    
 
    \item \textbf{1ES 1959+650 Region}: For the 1ES 1959+650 region, no zenith angle provides visibility from the CTAO South (see Table~\ref{tab:1ESvisibility}). In the North Array, the 40$^\circ$ zenith angle yields the longest visibility (1361 hours), but since improved observation quality is desired, the 20$^\circ$ IRF might be preferred if a reduction in visibility hours is acceptable. However, since only 40$^\circ$ and 60$^\circ$ IRFs provide visibility, the optimal choice remains the 40$^\circ$ IRF for this region, balancing visibility with a relatively lower zenith angle. Figure~\ref{fig:1ESarrays} illustrates the energy spectra for the W Comae region, both without (intrinsic) and with extragalactic background light (EBL) attenuation \cite{Saldana_Lopez_2021}. It also includes differential sensitivity curves optimized for different CTAO North configurations: the full array, achieving an integral sensitivity of $1.31 \times 10^{-12}\ \rm{cm^{-2}s^{-1}}$, CTAO North-MSTs ($1.38 \times 10^{-12}\ \rm{cm^{-2} s^{-1}}$), and CTAO North-LSTs ($1.62 \times 10^{-12}\ \rm{cm^{-2}s^{-1}}$). Additionally, Figure~\ref{fig:1ES_1959+650_region_sed} shows the SED, including observations from CTAO North and flux points of counterparts used in the likelihood analysis (see Table~\ref{tab:1ES}).
    
    \begin{table*}[h!]
    \begin{minipage}{\textwidth}
    \centering
    \begin{tabular}{lll}
    \toprule
     \textbf{Array} & \textbf{Zenith} & \textbf{Visibility} \\
    \midrule

    CTAO South& 20$^\circ$ & - \\
    CTAO South& 40$^\circ$ & - \\
    CTAO South& 60$^\circ$ & - \\
    CTAO North& 20$^\circ$ & - \\
    CTAO North& 40$^\circ$ & 1361.00 h \\
    CTAO North& 60$^\circ$ & 1193.50 h\\
    \bottomrule
    \end{tabular}
    \caption{Visibility of the 1ES 1959+650 region from the CTAO South and North arrays at different zenith angles. The calculations were conducted for 2025, considering typical nighttime hours (from 18:00 to 06:00 the next day) with 30-minute intervals. Visibility is expressed in hours, with dashes indicating that the source is not visible at the specified array and zenith angle.}\label{tab:1ESvisibility}
    \end{minipage}
    \end{table*}

    \begin{figure}[h!]
    \centering
    {\includegraphics[angle=0,width=0.8\textwidth]{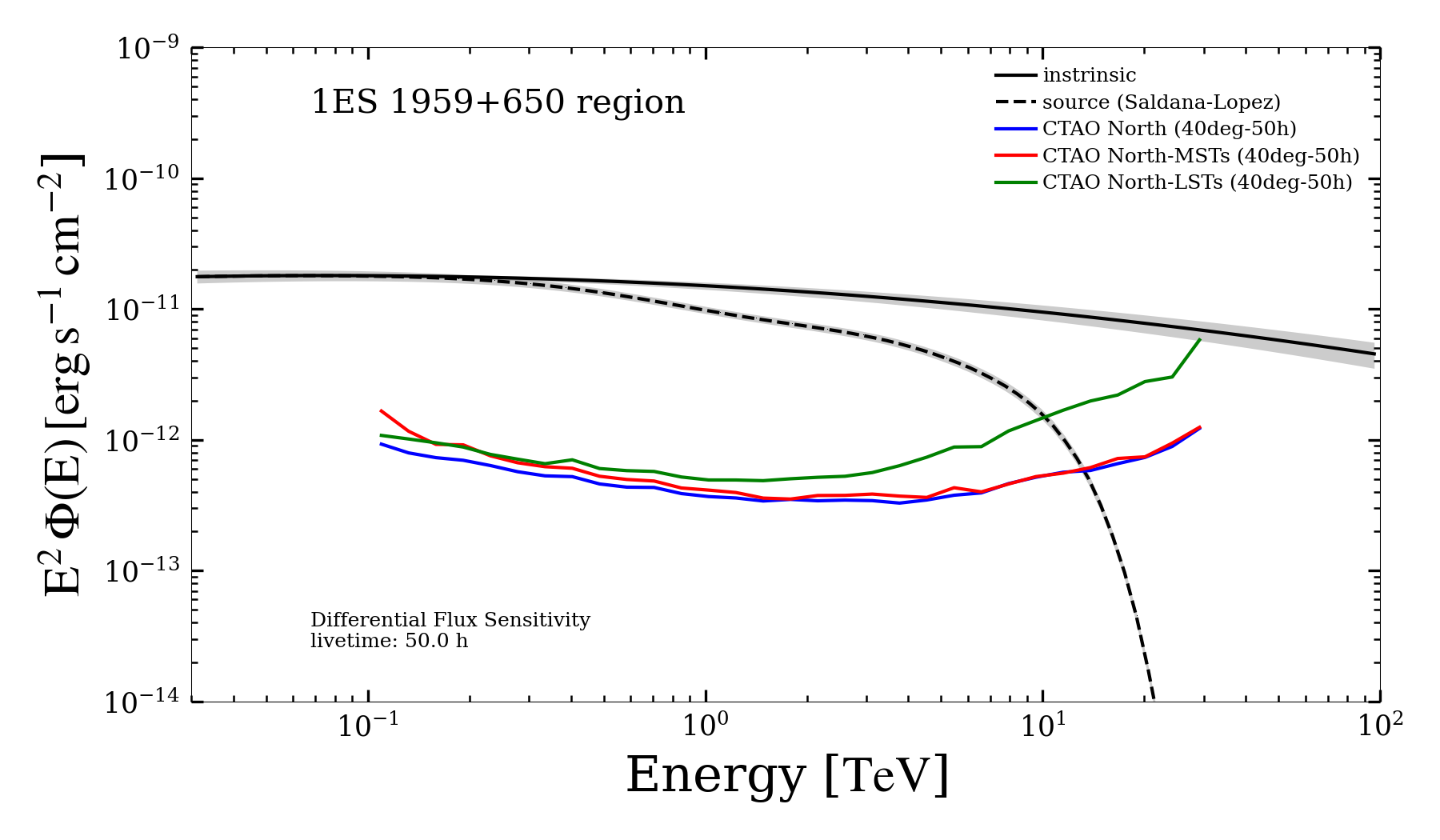}}
    \caption{
        Energy-dependent differential flux sensitivity and model predictions for the 1ES 1959+650 region. The black solid line represents the intrinsic model, while the black dashed line shows the absorbed spectrum model by Saldana-Lopez  \cite{Saldana_Lopez_2021} (see Table~\ref{tab:sourse_models}). The colored lines display sensitivity curves for the CTAO North with different array configurations: (blue line) CTAO North with 4 LSTs (Large-Sized Telescopes) and 9 MSTs (Medium-Sized Telescopes); (red line)  CTAO North-MSTs with 9 MSTs and (green line) CTAO North-LSTs with 4 LSTs. The IRFs are optimized to 40$^{\circ}$ zenith angle and the differential flux sensitivity refers to a 50-hour observation time.
    }
    \label{fig:1ESarrays}
    \end{figure}

    \begin{figure}[H]
    \centering
    {\includegraphics[angle=0,width=0.8\textwidth]{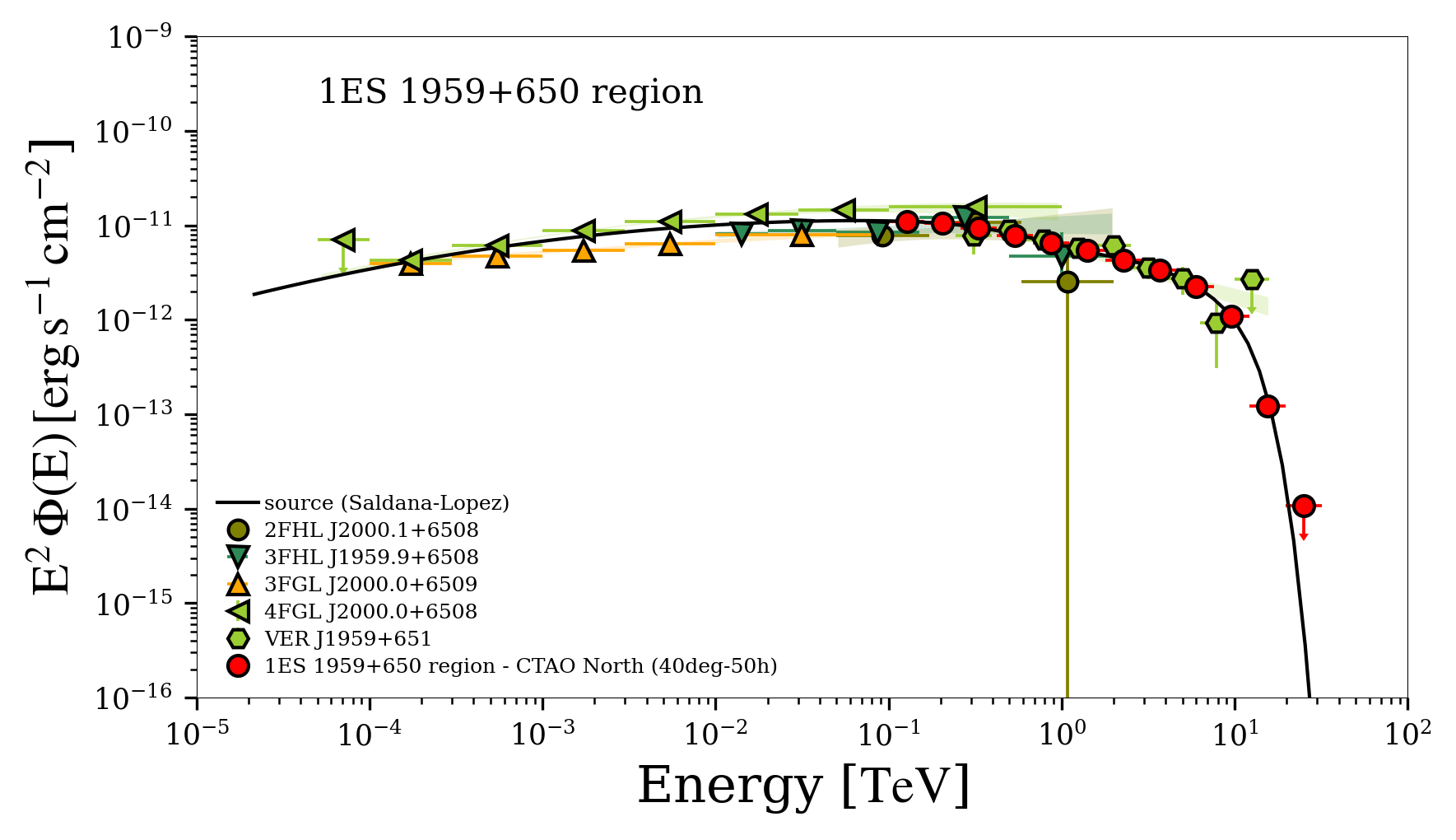}}
    \caption{
       Spectral energy distribution (SED) for the 1ES 1959+650 region. The plot shows the modeled gamma-ray flux for the source 1ES 1959+650 (black solid line), based on the model by Saldana-Lopez  \cite{Saldana_Lopez_2021} (see Table~\ref{tab:sourse_models}). Overlaid are observational data points from various gamma-ray catalogs: 2FHL, 3FHL, 3FGL, 4FGL, and VERITAS  (see Table~\ref{tab:pks2155}). The red circles represent the projected flux points for the 1ES 1959+650 region as observed with the CTAO North array for a 50-hour observation at a zenith angle of 40 degrees. Error bars indicate the statistical uncertainties for each data point.
    }
    \label{fig:1ES_1959+650_region_sed}
    \end{figure}
    
    \item \textbf{PKS 2005-489 Region}: Table~\ref{tab:PKS2005visibility} presents the visibility results for the PKS 2005-489 region. The CTAO South offers visibility at 20$^\circ$, 40$^\circ$, and 60$^\circ$ zenith angles. The 20$^\circ$ IRF provides 556 hours, which, although fewer than the 824 and 745 hours for 40$^\circ$ and 60$^\circ$ IRFs, is preferable due to better atmospheric conditions and observational quality. Therefore, the CTAO South 20$^\circ$ IRF is selected as the best option for this region. Figure~\ref{fig:PKS2005arrays} illustrates the energy spectra for the PKS 2005-489 region, both intrinsic and with extragalactic background light (EBL) attenuation applied, following \cite{Saldana_Lopez_2021} (see Table~\ref{tab:sourse_models}). The plot also shows differential sensitivity curves optimized for a zenith angle of 20$^\circ$ for various configurations of the CTAO South array: the full array with an integral sensitivity of $1.11 \times 10^{-12} \ \rm{cm^{-2} s^{-1}}$, CTAO South-MSTs with $1.36 \times 10^{-12} \ \rm{cm^{-2}  s^{-1}}$, and CTAO South-SSTs with $4.88 \times 10^{-13} \ \rm{cm^{-2}  s^{-1}}$. Although the CTAO South-SSTs exhibit lower integral sensitivity, it becomes apparent that their sensitivity is insufficient to fully characterize the spectrum. Figure~\ref{fig:PKS_2005-489_region_sed} presents the SED, including the CTAO South observations and the flux points of the counterparts used in the likelihood analysis (refer to Table~\ref{tab:pks2005}).
    
    \begin{table*}[h!]
    \begin{minipage}{\textwidth}
    \centering
    \begin{tabular}{lll}
    \toprule
     \textbf{Array} & \textbf{Zenith} & \textbf{Visibility} \\
    \midrule

    CTAO South& 20$^\circ$ & 556.00 h \\
    CTAO South& 40$^\circ$ & 824.00 h \\
    CTAO South& 60$^\circ$ & 745.00 h \\
    CTAO North& 20$^\circ$ & - \\
    CTAO North& 40$^\circ$ & -  \\
    CTAO North& 60$^\circ$ & - \\
    \bottomrule
    \end{tabular}
    \caption{Visibility of the PKS 2005-489 region from the CTAO South and North arrays at different zenith angles. Calculations were performed for 2025, accounting for typical nighttime hours (from 18:00 to 06:00 the following morning) with a 30-minute interval. Visibility is reported in hours, with dashes indicating instances where the source is not visible at the specified array and zenith angle.}\label{tab:PKS2005visibility}
    \end{minipage}
    \end{table*}
    
    \begin{figure}[h!]
    \centering
    {\includegraphics[angle=0,width=0.8\textwidth]{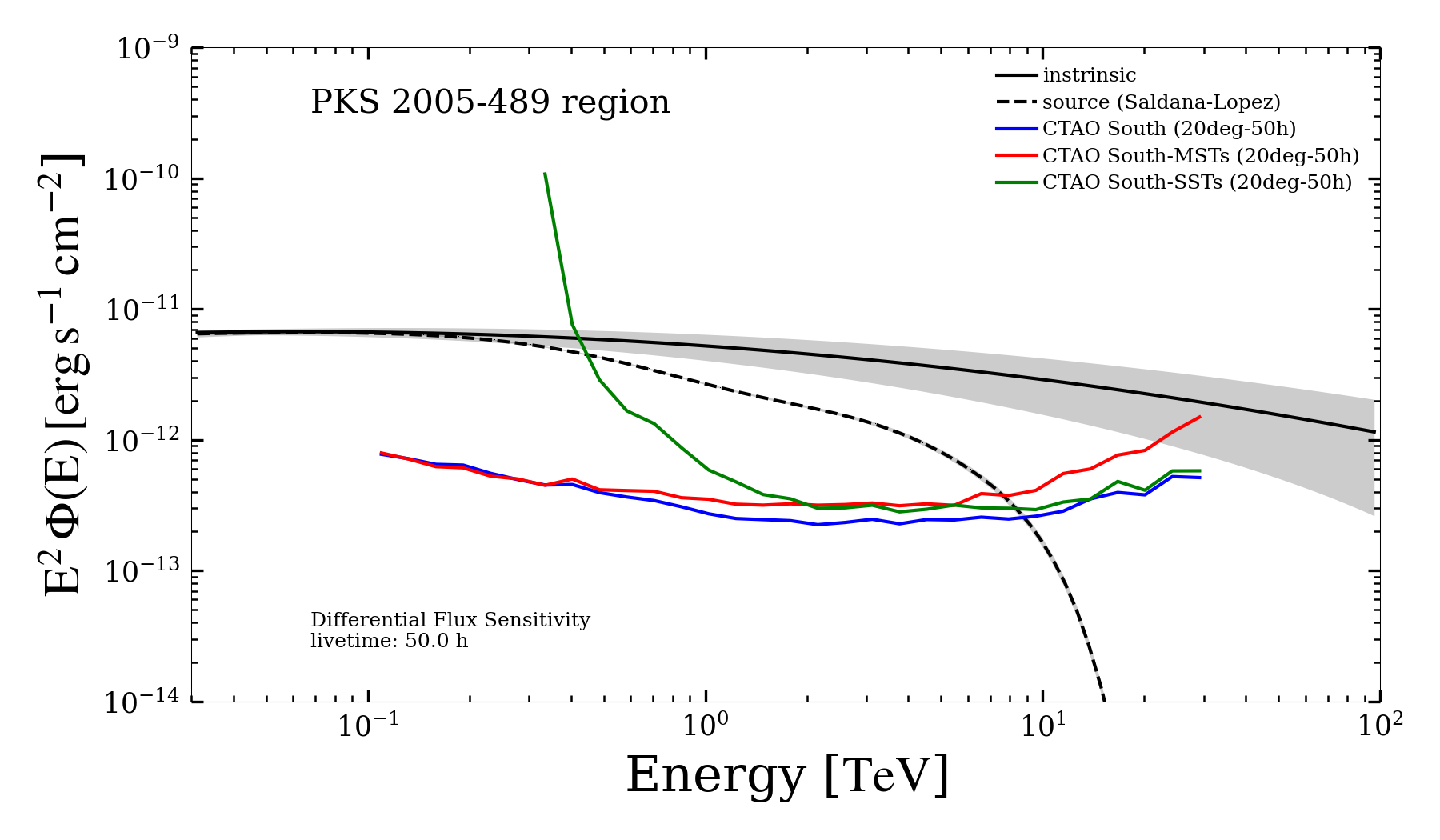}}
    \caption{
        Energy-dependent differential flux sensitivity and model predictions for the PKS 2005-489 region. The black solid line represents the intrinsic model, while the black dashed line shows the source model by Saldana-Lopez  \cite{Saldana_Lopez_2021} (see Table~\ref{tab:sourse_models}). The colored lines display sensitivity curves for the CTAO South with different array configurations: (blue line) CTAO South with 14 MSTs and 37 SSTs (Small-Sized Telescopes); (red line) CTAO South-MSTs with 14 MSTs and (green line) CTAO South-SSTs with 37 SSTs. The IRFs are optimized to 20$^{\circ}$ zenith angle and the differential flux sensitivity refers to a 50-hour observation time.
    }
    \label{fig:PKS2005arrays}
    \end{figure}

\begin{figure}[H]
    \centering
    {\includegraphics[angle=0,width=0.8\textwidth]{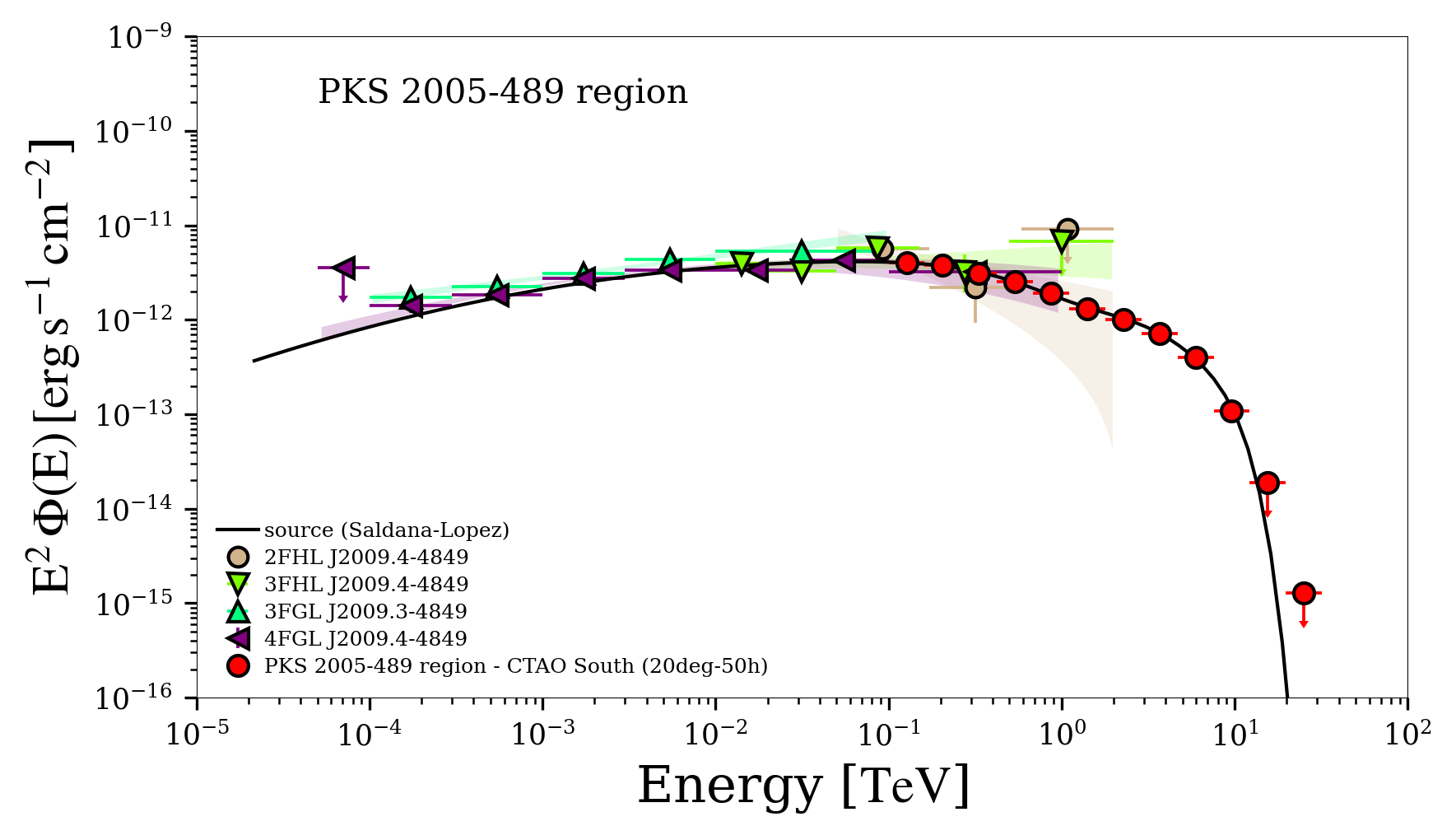}}
    \caption{
    Spectral energy distribution (SED) for the PKS 2005-489 region. This plot shows the modeled gamma-ray flux for PKS 2005-489 based on the absorbed spectrum model from Saldana-Lopez, represented by the black solid line \cite{Saldana_Lopez_2021} (see Table~\ref{tab:sourse_models}). Overlaid are the observational data points from various Fermi catalogs: 2FHL, 3FHL, 3FGL, and 4FGL, each denoted with distinct markers and colors. The red circles indicate the projected sensitivity of the PKS 2005-489 region observed with the CTAO South array at a zenith angle of 20 degrees for a 50-hour observation. Error bars show the statistical uncertainties for each flux point.
    }
    \label{fig:PKS_2005-489_region_sed}
    \end{figure}
    
    \item \textbf{PKS 2155-304 Region}:
    For the PKS 2155-304 region, the CTAO South’s 20$^\circ$ zenith angle offers 580.5 hours (see Table~\ref{tab:PKS2155_visibility}), making it the optimal choice due to the lower zenith angle and comparable visibility hours. The CTAO North provides visibility only at 60$^\circ$ (967 hours), making it the sole option for observing this region from the North. Thus, the CTAO South (as indicated in \cite{CTACollaboration} - Table 12.1) and the 20$^\circ$ zenith angle are recommended for superior observational quality. Figure~\ref{fig:PKS2155arrays} displays energy spectra without (intrinsic) and with EBL attenuation \cite{Saldana_Lopez_2021} (see Table~\ref{tab:sourse_models}), as well as differential sensitivity optimized for a zenith angle of 20$^\circ$ curves of CTAO South with integral sensitivity of $1.11 \times 10^{-12} \rm{cm^{-1}} s^{-1}$, CTAO South-MSTs ($1.36 \times 10^{-12} \rm{cm^{-1}} s^{-1}$), and CTAO South-SSTs ($4.88 \times 10^{-13} \rm{cm^{-1}} s^{-1}$). Although the CTAO South-SSTs have a lower sensitivity integral, it is evident that the sensitivity is insufficient for assessing the spectrum as a whole. Figure~\ref{fig:PKS_2155-304_region_sed} displays the SED, CTAO South observations, and flux points of the counterparts used in the likelihood analysis (see Table~\ref{tab:pks2155}).
    
    \begin{table*}[h!]
    \begin{minipage}{\textwidth}
    \centering
    \begin{tabular}{lll}
    \toprule
     \textbf{Array} & \textbf{Zenith} & \textbf{Visibility} \\
    \midrule
    
    CTAO South& 20$^\circ$ & 580.50 h \\
    CTAO South& 40$^\circ$ & 564.00  h \\
    CTAO South& 60$^\circ$ & 578.50 h \\
    CTAO North& 20$^\circ$ & - \\
    CTAO North& 40$^\circ$ & -  \\
    CTAO North& 60$^\circ$ & 967.00 h\\
    \bottomrule
    \end{tabular}
    \caption{Visibility of the PKS 2155-304 region from the CTAO South and North arrays at different zenith angles. Calculations are based on 2025 with a time step of 30 minutes. Visibility is given in hours and the ashes indicating the source is not visible at the corresponding array and zenith angle.}\label{tab:PKS2155_visibility}
    \end{minipage}
    \end{table*}

    \begin{figure}[h!]
    \centering
    {\includegraphics[angle=0,width=0.8\textwidth]{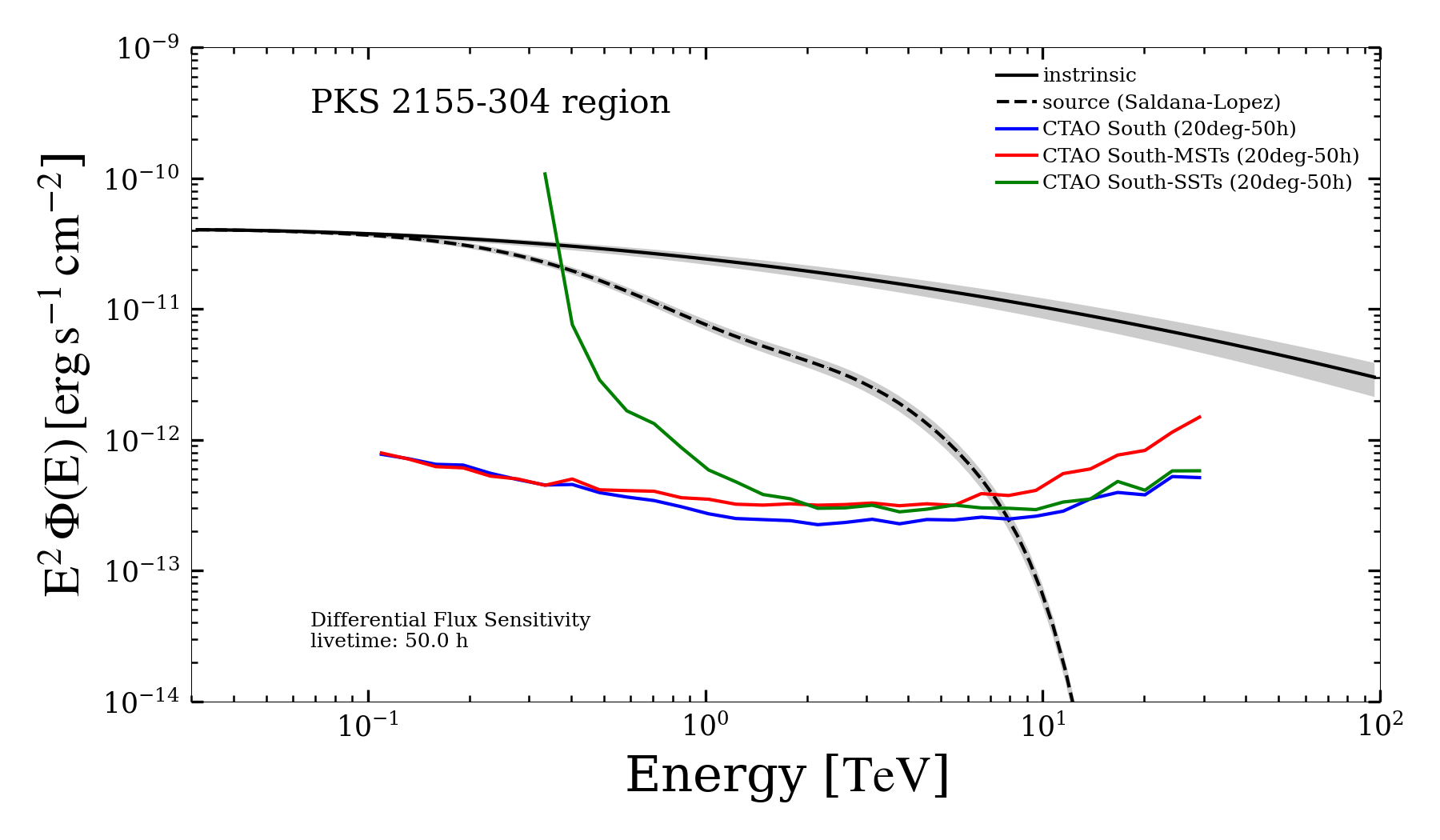}}
    \caption{
     Energy-dependent differential flux sensitivity and spectral model for the PKS 2155-304 region. The black solid line represents the intrinsic model, while the black dashed line shows the absorbed spectrum model by Saldana-Lopez  \cite{Saldana_Lopez_2021} (see Table~\ref{tab:sourse_models}). The colored lines display sensitivity curves for the CTAO South with different array configurations: (blue line) CTAO South with 14 MSTs and 37 SSTs (Small-Sized Telescopes); (red line) CTAO South-MSTs with 14 MSTs and (green line) CTAO South-SSTs with 37 SSTs. The IRFs are optimized to 20$^{\circ}$ zenith angle and the differential flux sensitivity refers to a 50-hour observation time.}\label{fig:PKS2155arrays}
    \end{figure}

        \begin{figure}[H]
    \centering
    {\includegraphics[angle=0,width=0.8\textwidth]{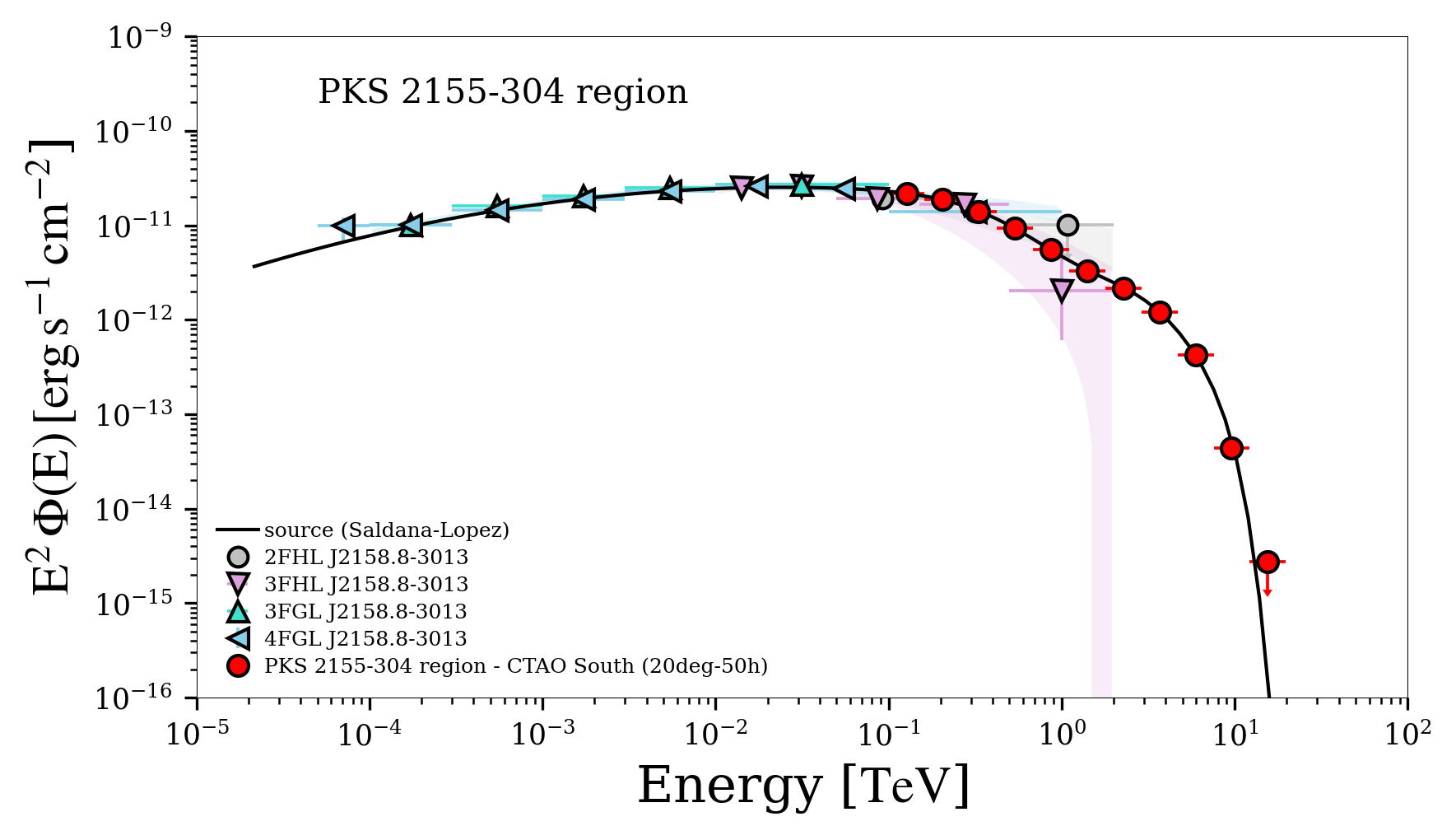}}
    \caption{
        Spectral energy distribution (SED) for the PKS 2155-304 region. This plot displays the modeled gamma-ray flux for PKS 2155-304, overlaid with observational data from Fermi-LAT catalogs: 3FHL, 3FGL, and 4FGL (see Table~\ref{tab:pks2155}). The black solid line represents the source model by Saldana-Lopez \cite{Saldana_Lopez_2021} (see Table~\ref{tab:sourse_models}) and the red circles show the flux points from the CTAO South array for a 50-hour observation. Error bars indicate the statistical uncertainties for each data point.
            }
    \label{fig:PKS_2155-304_region_sed}
    \end{figure}
\end{itemize}

Table~\ref{tab:CTAOspectral_models} summarizes the spectral model parameters derived from CTAO observations for four blazar regions: W Comae, 1ES 1959+650, PKS 2005-489, and PKS 2155-304. The parameters include the normalization factor $\Phi_0$, spectral index $\alpha$, curvature parameter $\beta$, and cutoff energy $E_{\rm cut}$, each relevant to characterizing the energy spectrum of these sources.

\begingroup
\renewcommand*{\thefootnote}{\alph{footnote}}  
\begin{table}[htbp]
\centering
\begin{minipage}{\textwidth}
    \centering
\begin{tabular}{ccccc}
\toprule
\textbf{Region}	& \textbf{W Comae} & \textbf{1ES 1959+650} & \textbf{PKS 2005-489} & \textbf{PKS 2155-304} \\
\midrule

$\Phi_0 \times 10^{-13}$ & 
$20.418 \pm 1.241$ & $783.560 \pm 10.620$ & $176.760 \pm 7.400$ & $508.900 \pm 4.404$ \\
$(\mathrm{cm^{-2}\,s^{-1}\,TeV^{-1}})$\\

$\alpha$ & 
$3.209 \pm 0.086$ & $2.188 \pm 0.008$ & $2.682 \pm 0.013$ & $3.152 \pm 0.010$ \\

$\beta \times 10^{-2}$ & 
$17.04 \pm 4.04$ & - & $13.77 \pm 0.76$ & $21.51 \pm 0.46$ \\

$\mathrm{E_{\rm cut}}$ (TeV) & 
- & $5.41 \pm 0.19$ & - & - \\

\bottomrule
\end{tabular}
\caption{Spectral model parameters, amplitude $\Phi_0$ (scaled by $10^{-13}$),  spectral index $\alpha$ and curvature parameter $\beta$ (scaled by $10^{-2}$) and cutoff energy $E_{\rm cut}$ of the CTAO observations 
for the four blazar regions—W Comae, 1ES 1959+650, PKS 2005-489, and PKS 2155-304 (see Equations~\ref{eq:LP} and \ref{eq:ECPL}).}
\label{tab:CTAOspectral_models}
\end{minipage}
\end{table}
\endgroup

\section{Conclusions}\label{sec:conclusions}

This study explored the potential of four BL Lac objects (W Comae, 1ES 1959+650, PKS 2005-489, and PKS 2155-304) as possible sources of astrophysical neutrinos, gamma rays, and cosmic rays.  Using a lepto-hadronic model and building on the findings of Rodrigues et al. (2024) \cite{2024A&A...681A.119R}, the research analyzed the interactions between high-energy protons and ambient photons within blazar jets to investigate neutrino production and gamma-ray emission from charged particles. The results have indicated that a purely leptonic model is insufficient to explain the high-energy gamma-ray emissions observed in W Comae and 1ES 1959+650. The inclusion of hadronic interactions (pp and p$\gamma$) significantly improves the fit of the model for these two BL Lacs, suggesting a substantial contribution of hadronic processes to their high-energy emission. In contrast, for PKS 2005-489 and PKS 2155-304, the leptonic model provides a reasonably good fit, although there is a suggestion of hadronic contributions at high energies. In conclusion, the lepto-hadronic model successfully describes the multiwavelength observations throughout a wide energy range for all four sources, highlighting the need for considering hadronic processes to provide a complete understanding of BL Lac emission.

The final section evaluated the Cherenkov Telescope Array Observatory's (CTAO) ability to detect gamma-ray emissions from the four BL Lacs.  A likelihood analysis incorporating multiwavelength observational data (multi-GeV to multi-TeV) demonstrated CTAO's detection capabilities.  However, the effectiveness varied considerably depending on the target BL Lac and specific CTAO array configuration (North or South) and zenith angle. For W Comae, only the CTAO South array at the zenith $60^{\circ}$ offers visibility (1263 hours), despite the increased atmospheric absorption, while the North array provides superior viewing at $20^{\circ}$ (556.5 hours). The 1ES 1959+650 source is only visible from the CTAO North array (best at $40^{\circ}$, 1361 hours), although $20^{\circ}$ is preferred for quality. PKS 2005-489 shows visibility from the CTAO South at all zenith angles (best at $40^{\circ}$, 824 hours), with $20^{\circ}$ favored for observation quality. Finally, PKS 2155-304 is best observed from the CTAO South at $20^{\circ}$ (580.5 hours), while the North array only provides visibility at $60^{\circ}$ (967 hours). In summary, the CTAO North array generally offers superior visibility and observational quality compared to the South array, although the optimal zenith angle and array choice depend on the specific BL Lac. Furthermore, while the SED for 1ES 1959+650 extended to higher energies ($\sim$ 30 TeV) than the other BL Lacs,  all four sources demonstrated a potential for CTAO observations exceeding 10 TeV. Future observations at multi-TeV energies by the CTAO will have the ability to detect gamma-ray emissions from the blazars discussed here. The results suggest that these sources could be effective cosmic-ray emitters. However, further multi-messenger observations are crucial to confirm this conclusion. The CTAO will play a significant role in these future observations, enhancing our understanding of the origins of multimessenger particles and the role of blazars in their production.

\bmhead{Acknowledgements}
We express our gratitude to the anonymous referee for providing thoughtful and constructive feedback, whose input has greatly enhanced the quality of this work. The authors acknowledge the AWS Cloud Credit/CNPq and the National Laboratory for Scientific Computing (LNCC/MCTI, Brazil) for providing HPC resources of the SDumont supercomputer, which have contributed to the research results reported in this paper. URL: https://sdumont.lncc.br. This study used gamma-cat, an open data repository and the source catalog for gamma-ray astronomy, available at \href{https://github.com/gammapy/gamma-cat}{https://github.com/gammapy/gamma-cat}. The research employed the ATNF Pulsar Catalogue \cite{Manchester2005}, accessible at \href{https://www.atnf.csiro.au/research/pulsar/psrcat/}{https://www.atnf.csiro.au/research/pulsar/psrcat/}. The research also employed Gammapy, a Python package developed by the community for TeV gamma-ray astronomy \citep{Deil2017}, accessible at \href{https://www.gammapy.org}{https://www.gammapy.org}. In addition, we used the instrument response functions for the Cherenkov Telescope Array Observatory (CTAO) provided by the CTA Consortium and CTAO. For detailed information on these instrument response functions, see \href{https://www.ctao-observatory.org/science/cta-performance}{https://www.ctao-observatory.org/science/cta-performance} (version prod5 v0.1; \cite{CTAOIRFS}).

\section*{Declarations}

\begin{itemize}

\item Funding

R.S. acknowledges financial support from the Coordenação de Aperfeiçoamento de Pessoal de Nível Superior – Brasil (CAPES) – Finance Code 001. R.C.A., R.J.C. and R.S. acknowledge the financial support from the NAPI “Fenômenos Extremos do Universo” of Fundação de Apoio à Ciência, Tecnologia e Inovação do Paraná. L.A.S.P. gratefully acknowledges financial support from FAPESP under grant number 2024/14769-9. The research of R.C.A is supported by the CAPES/Alexander von Humboldt Program (88881.800216/2022-01), Conselho Nacional de Desenvolvimento Cient\'{i}fico e Tecnol\'{o}gico (CNPq) grant numbers (307750/2017-5) and (401634/2018-3/AWS), Araucária Foundation (698/2022) and (721/2022) and FAPESP (2021/01089-1). She also thanks L'Oreal Brazil for the support, with the partnership of ABC and UNESCO in Brazil.

\item Conflict of interest

The authors declare no competing interests

\item Ethics approval and consent to participate

\item Consent for publication

\item Data availability 

No datasets were generated during the current study.

\item Materials availability

\item Author contribution

All authors contributed equally.

\end{itemize}

\bibliography{sn-bibliography}

\begin{thebibliography}{10}
\providecommand{\doi}[1]{\url{https://doi.org/#1}}
\bibcommenthead

\bibitem[\protect\citeauthoryear{{Aab} et~al.}{2018}]{2018ApJ...853L..29A}
{Aab} A, {Abreu} P, {Aglietta} M, {Albuquerque} IFM, {Allekotte} I, {Almela} A, et~al.
\newblock {An Indication of Anisotropy in Arrival Directions of Ultra-high-energy Cosmic Rays through Comparison to the Flux Pattern of Extragalactic Gamma-Ray Sources}.
\newblock \apjl. 2018 Feb;853(2):L29.
\newblock \doi{10.3847/2041-8213/aaa66d}.
\newblock {\href{https://arxiv.org/abs/1801.06160}{{arXiv:1801.06160}}}. {[astro-ph.HE]}.

\bibitem[\protect\citeauthoryear{Fermi}{1949}]{PhysRev.75.1169}
Fermi E.
\newblock On the Origin of the Cosmic Radiation.
\newblock Phys Rev. 1949 Apr;75:1169--1174.
\newblock \doi{10.1103/PhysRev.75.1169}.

\bibitem[\protect\citeauthoryear{{Pereira} et~al.}{2024}]{2024PhRvL.132i1401P}
{Pereira} JP, {Coimbra-Ara{\'u}jo} CH, {dos Anjos} RC, {Coelho} JG.
\newblock {Binary Coalescences as Sources of Ultrahigh-Energy Cosmic Rays}.
\newblock \prl. 2024 Feb;132(9):091401.
\newblock \doi{10.1103/PhysRevLett.132.091401}.
\newblock {\href{https://arxiv.org/abs/2307.06200}{{arXiv:2307.06200}}}. {[astro-ph.HE]}.

\bibitem[\protect\citeauthoryear{{Murase} et~al.}{2012}]{2012ApJ...749...63M}
{Murase} K, {Dermer} CD, {Takami} H, {Migliori} G.
\newblock {Blazars as Ultra-high-energy Cosmic-ray Sources: Implications for TeV Gamma-Ray Observations}.
\newblock \apj. 2012 Apr;749(1):63.
\newblock \doi{10.1088/0004-637X/749/1/63}.
\newblock {\href{https://arxiv.org/abs/1107.5576}{{arXiv:1107.5576}}}. {[astro-ph.HE]}.

\bibitem[\protect\citeauthoryear{{Zhang} et~al.}{2014}]{2014IJAA....4..499Z}
{Zhang} B, {Zhao} X, {Cao} Z.
\newblock {TeV Blazars as the Sources of Ultra High Energy Cosmic Rays}.
\newblock International Journal of Astronomy and Astrophysics. 2014 Jan;4(3):499--509.
\newblock \doi{10.4236/ijaa.2014.43046}.

\bibitem[\protect\citeauthoryear{{Hillas}}{1984}]{1984ARA&A..22..425H}
{Hillas} AM.
\newblock {The Origin of Ultra-High-Energy Cosmic Rays}.
\newblock \araa. 1984 Jan;22:425--444.
\newblock \doi{10.1146/annurev.aa.22.090184.002233}.

\bibitem[\protect\citeauthoryear{{Zhang} et~al.}{2014}]{Zhang2014}
{Zhang} B, {Zhao} X, {Cao} Z.
\newblock {TeV Blazars as the Sources of Ultra High Energy Cosmic Rays}.
\newblock International Journal of Astronomy and Astrophysics. 2014 Jan;4(3):499--509.
\newblock \doi{10.4236/ijaa.2014.43046}.

\bibitem[\protect\citeauthoryear{{Das} et~al.}{2022}]{Das2022}
{Das} S, {Gupta} N, {Razzaque} S.
\newblock {Implications of multiwavelength spectrum on cosmic-ray acceleration in blazar TXS 0506+056}.
\newblock \aap. 2022 Dec;668:A146.
\newblock \doi{10.1051/0004-6361/202244653}.
\newblock {\href{https://arxiv.org/abs/2208.00838}{{arXiv:2208.00838}}}. {[astro-ph.HE]}.

\bibitem[\protect\citeauthoryear{{Sigl}}{2009}]{2009NJPh...11f5014S}
{Sigl} G.
\newblock {Time structure and multi-messenger signatures of ultra-high energy cosmic ray sources}.
\newblock New Journal of Physics. 2009 Jun;11(6):065014.
\newblock \doi{10.1088/1367-2630/11/6/065014}.
\newblock {\href{https://arxiv.org/abs/0803.3800}{{arXiv:0803.3800}}}. {[astro-ph]}.

\bibitem[\protect\citeauthoryear{{Unger} and {Farrar}}{2023}]{2023arXiv231112120U}
{Unger} M, {Farrar} GR.
\newblock {The Coherent Magnetic Field of the Milky Way}.
\newblock arXiv e-prints. 2023 Nov;p. arXiv:2311.12120.
\newblock \doi{10.48550/arXiv.2311.12120}.
\newblock {\href{https://arxiv.org/abs/2311.12120}{{arXiv:2311.12120}}}. {[astro-ph.GA]}.

\bibitem[\protect\citeauthoryear{{Globus} and {Blandford}}{2023}]{2023EPJWC.28304001G}
{Globus} N, {Blandford} R.
\newblock {Ultra High Energy Cosmic Ray Source Models: Successes, Challenges and General Predictions}.
\newblock In: European Physical Journal Web of Conferences. vol. 283 of European Physical Journal Web of Conferences; 2023. p. 04001.

\bibitem[\protect\citeauthoryear{{Albert} et~al.}{2022}]{2022ApJ...934..164A}
{Albert} A, {Alves} S, {Andr{\'e}} M, {Anghinolfi} M, {A } M, {Ardid} S, et~al.
\newblock {Search for Spatial Correlations of Neutrinos with Ultra-high-energy Cosmic Rays}.
\newblock \apj. 2022 Aug;934(2):164.
\newblock \doi{10.3847/1538-4357/ac6def}.
\newblock {\href{https://arxiv.org/abs/2201.07313}{{arXiv:2201.07313}}}. {[astro-ph.HE]}.

\bibitem[\protect\citeauthoryear{{Sharma}}{2024}]{2024Univ...10..326S}
{Sharma} A.
\newblock {Multi-Messenger Connection in High-Energy Neutrino Astronomy}.
\newblock Universe. 2024 Aug;10(8):326.
\newblock \doi{10.3390/universe10080326}.
\newblock {\href{https://arxiv.org/abs/2408.11353}{{arXiv:2408.11353}}}. {[astro-ph.HE]}.

\bibitem[\protect\citeauthoryear{{Giommi} and {Padovani}}{2021}]{2021Univ....7..492G}
{Giommi} P, {Padovani} P.
\newblock {Astrophysical Neutrinos and Blazars}.
\newblock Universe. 2021 Dec;7(12):492.
\newblock \doi{10.3390/universe7120492}.
\newblock {\href{https://arxiv.org/abs/2112.06232}{{arXiv:2112.06232}}}. {[astro-ph.HE]}.

\bibitem[\protect\citeauthoryear{Collaboration: and collaborators:}{2024}]{MAGIC2024}
Collaboration: M, collaborators: M.
\newblock {The variability patterns of the TeV blazar PG 1553 + 113 from a decade of MAGIC and multiband observations}.
\newblock Monthly Notices of the Royal Astronomical Society. 2024 03;529(4):3894--3911.
\newblock \doi{10.1093/mnras/stae649}.
\newblock {\href{https://arxiv.org/abs/https://academic.oup.com/mnras/article-pdf/529/4/3894/58613552/stae649.pdf}{{https://academic.oup.com/mnras/article-pdf/529/4/3894/58613552/stae649.pdf}}}.

\bibitem[\protect\citeauthoryear{{Murase} and {Stecker}}{2023}]{2023ecnp.book..483M}
{Murase} K, {Stecker} FW.
\newblock {High-Energy Neutrinos from Active Galactic Nuclei}.
\newblock In: {Stecker} FW, editor. The Encyclopedia of Cosmology. Set 2: Frontiers in Cosmology. Volume 2: Neutrino Physics and Astrophysics; 2023. p. 483--540.

\bibitem[\protect\citeauthoryear{{Mannheim} et~al.}{1992}]{1992A&A...260L...1M}
{Mannheim} K, {Stanev} T, {Biermann} PL.
\newblock {Neutrinos from flat-spectrum radio quasars}.
\newblock \aap. 1992 Jul;260(1-2):L1--L3.

\bibitem[\protect\citeauthoryear{Aartsen et~al.}{2017}]{Aartsen_2017}
Aartsen MG, Ackermann M, Adams J, Aguilar JA, Ahlers M, Ahrens M, et~al.
\newblock The IceCube Neutrino Observatory: instrumentation and online systems.
\newblock Journal of Instrumentation. 2017 mar;12(03):P03012.
\newblock \doi{10.1088/1748-0221/12/03/P03012}.

\bibitem[\protect\citeauthoryear{Collaboration}{2018}]{2018Sci...361.1378I}
Collaboration I.
\newblock {Multimessenger observations of a flaring blazar coincident with high-energy neutrino IceCube-170922A}.
\newblock Science. 2018 Jul;361(6398):eaat1378.
\newblock \doi{10.1126/science.aat1378}.
\newblock {\href{https://arxiv.org/abs/1807.08816}{{arXiv:1807.08816}}}. {[astro-ph.HE]}.

\bibitem[\protect\citeauthoryear{Collaboration}{2018}]{2018Sci...361..147I}
Collaboration I.
\newblock {Neutrino emission from the direction of the blazar TXS 0506+056 prior to the IceCube-170922A alert}.
\newblock Science. 2018 Jul;361(6398):147--151.
\newblock \doi{10.1126/science.aat2890}.
\newblock {\href{https://arxiv.org/abs/1807.08794}{{arXiv:1807.08794}}}. {[astro-ph.HE]}.

\bibitem[\protect\citeauthoryear{{Das} et~al.}{2020}]{2020ApJ...889..149D}
{Das} S, {Gupta} N, {Razzaque} S.
\newblock {Ultrahigh-energy Cosmic-Ray Interactions as the Origin of Very High-energy {\ensuremath{\gamma}}-Rays from BL Lacertae Objects}.
\newblock \apj. 2020 Feb;889(2):149.
\newblock \doi{10.3847/1538-4357/ab6131}.
\newblock {\href{https://arxiv.org/abs/1911.06011}{{arXiv:1911.06011}}}. {[astro-ph.HE]}.

\bibitem[\protect\citeauthoryear{{Das} et~al.}{2022}]{2022A&A...658L...6D}
{Das} S, {Razzaque} S, {Gupta} N.
\newblock {Cosmogenic gamma-ray and neutrino fluxes from blazars associated with IceCube events}.
\newblock \aap. 2022 Feb;658:L6.
\newblock \doi{10.1051/0004-6361/202142123}.
\newblock {\href{https://arxiv.org/abs/2108.12120}{{arXiv:2108.12120}}}. {[astro-ph.HE]}.

\bibitem[\protect\citeauthoryear{Healey et~al.}{2008}]{Healey2008}
Healey SE, Romani RW, Cotter G, Michelson PF, Schlafly EF, Readhead ACS, et~al.
\newblock CGRaBS: An All-Sky Survey of Gamma-Ray Blazar Candidates.
\newblock The Astrophysical Journal Supplement Series. 2008 mar;175(1):97.
\newblock \doi{10.1086/523302}.

\bibitem[\protect\citeauthoryear{{Rodrigues} et~al.}{2024}]{2024A&A...681A.119R}
{Rodrigues} X, {Paliya} VS, {Garrappa} S, {Omeliukh} A, {Franckowiak} A, {Winter} W.
\newblock {Leptohadronic multi-messenger modeling of 324 gamma-ray blazars}.
\newblock \aap. 2024 jan;681:A119.
\newblock \doi{10.1051/0004-6361/202347540}.
\newblock {\href{https://arxiv.org/abs/2307.13024}{{arXiv:2307.13024}}}. {[astro-ph.HE]}.

\bibitem[\protect\citeauthoryear{{Greiner} et~al.}{2008}]{Greiner2008}
{Greiner} J, {Bornemann} W, {Clemens} C, {Deuter} M, {Hasinger} G, {Honsberg} M, et~al.
\newblock {GROND{\textemdash}a 7-Channel Imager}.
\newblock \pasp. 2008 Apr;120(866):405.
\newblock \doi{10.1086/587032}.
\newblock {\href{https://arxiv.org/abs/0801.4801}{{arXiv:0801.4801}}}. {[astro-ph]}.

\bibitem[\protect\citeauthoryear{{Roming} et~al.}{2005}]{Roming2005}
{Roming} PWA, {Kennedy} TE, {Mason} KO, {Nousek} JA, {Ahr} L, {Bingham} RE, et~al.
\newblock {The Swift Ultra-Violet/Optical Telescope}.
\newblock \ssr. 2005 Oct;120(3-4):95--142.
\newblock \doi{10.1007/s11214-005-5095-4}.
\newblock {\href{https://arxiv.org/abs/astro-ph/0507413}{{arXiv:astro-ph/0507413}}}. {[astro-ph]}.

\bibitem[\protect\citeauthoryear{Harrison et~al.}{2013}]{Harrison2013}
Harrison FA, Craig WW, Christensen FE, Hailey CJ, Zhang WW, Boggs SE, et~al.
\newblock THE NUCLEAR SPECTROSCOPIC TELESCOPE ARRAY (NuSTAR) HIGH-ENERGY X-RAY MISSION.
\newblock The Astrophysical Journal. 2013 may;770(2):103.
\newblock \doi{10.1088/0004-637X/770/2/103}.

\bibitem[\protect\citeauthoryear{{Matthews} et~al.}{2001}]{Matthews2001}
{Matthews} LD, {van Driel} W, {Monnier-Ragaigne} D.
\newblock {H I observations of giant low surface brightness galaxies}.
\newblock \aap. 2001 Jan;365:1--10.
\newblock \doi{10.1051/0004-6361:20000002}.
\newblock {\href{https://arxiv.org/abs/astro-ph/0010075}{{arXiv:astro-ph/0010075}}}. {[astro-ph]}.

\bibitem[\protect\citeauthoryear{Weisskopf et~al.}{2000}]{Weisskopf2000}
Weisskopf MC, Tananbaum HD, van Speybroeck LP, O'Dell SL.
\newblock {Chandra x-ray observatory (cxo):overview}.
\newblock Proc SPIE Int Soc Opt Eng. 2000;4012:2.
\newblock \doi{10.1117/12.391545}.
\newblock {\href{https://arxiv.org/abs/astro-ph/0004127}{{arXiv:astro-ph/0004127}}}.

\bibitem[\protect\citeauthoryear{{Burrows} et~al.}{2005}]{Burrows2005}
{Burrows} DN, {Hill} JE, {Nousek} JA, {Kennea} JA, {Wells} A, {Osborne} JP, et~al.
\newblock {The Swift X-Ray Telescope}.
\newblock \ssr. 2005 Oct;120(3-4):165--195.
\newblock \doi{10.1007/s11214-005-5097-2}.
\newblock {\href{https://arxiv.org/abs/astro-ph/0508071}{{arXiv:astro-ph/0508071}}}. {[astro-ph]}.

\bibitem[\protect\citeauthoryear{{Krimm} et~al.}{2013}]{Krimm2013}
{Krimm} HA, {Holland} ST, {Corbet} RHD, {Pearlman} AB, {Romano} P, {Kennea} JA, et~al.
\newblock {The Swift/BAT Hard X-Ray Transient Monitor}.
\newblock \apjs. 2013 Nov;209(1):14.
\newblock \doi{10.1088/0067-0049/209/1/14}.
\newblock {\href{https://arxiv.org/abs/1309.0755}{{arXiv:1309.0755}}}. {[astro-ph.HE]}.

\bibitem[\protect\citeauthoryear{Atwood et~al.}{2009}]{Atwood2009}
Atwood WB, Abdo AA, Ackermann M, Althouse W, Anderson B, Axelsson M, et~al.
\newblock THE LARGE AREA TELESCOPE ON THE FERMI GAMMA-RAY SPACE TELESCOPE MISSION.
\newblock The Astrophysical Journal. 2009 may;697(2):1071.
\newblock \doi{10.1088/0004-637X/697/2/1071}.

\bibitem[\protect\citeauthoryear{{Paliya} et~al.}{2017}]{2017ApJ...851...33P}
{Paliya} VS, {Marcotulli} L, {Ajello} M, {Joshi} M, {Sahayanathan} S, {Rao} AR, et~al.
\newblock {General Physical Properties of CGRaBS Blazars}.
\newblock \apjl. 2017 dec;851(1):33.
\newblock \doi{10.3847/1538-4357/aa98e1}.
\newblock {\href{https://arxiv.org/abs/1711.01292}{{arXiv:1711.01292}}}. {[astro-ph.HE]}.

\bibitem[\protect\citeauthoryear{{Acciari} et~al.}{2008}]{2008ApJ...684L..73A}
{Acciari} VA, {Aliu} E, {Beilicke} M, {Benbow} W, {B{\"o}ttcher} M, {Bradbury} SM, et~al.
\newblock {VERITAS Discovery of 200 GeV Gamma-Ray Emission from the Intermediate-Frequency-Peaked BL Lacertae Object W Comae}.
\newblock \apjl. 2008 Sep;684(2):L73.
\newblock \doi{10.1086/592244}.
\newblock {\href{https://arxiv.org/abs/0808.0889}{{arXiv:0808.0889}}}. {[astro-ph]}.

\bibitem[\protect\citeauthoryear{{Acciari} et~al.}{2009}]{2009ApJ...707..612A}
{Acciari} VA, {Aliu} E, {Aune} T, {Beilicke} M, {Benbow} W, {B{\"o}ttcher} M, et~al.
\newblock {Multiwavelength Observations of a TeV-Flare from W Comae}.
\newblock \apj. 2009 Dec;707(1):612--620.
\newblock \doi{10.1088/0004-637X/707/1/612}.
\newblock {\href{https://arxiv.org/abs/0910.3750}{{arXiv:0910.3750}}}. {[astro-ph.HE]}.

\bibitem[\protect\citeauthoryear{{Krawczynski} et~al.}{2004}]{2004ApJ...601..151K}
{Krawczynski} H, {Hughes} SB, {Horan} D, {Aharonian} F, {Aller} MF, {Aller} H, et~al.
\newblock {Multiwavelength Observations of Strong Flares from the TeV Blazar 1ES 1959+650}.
\newblock \apj. 2004 Jan;601(1):151--164.
\newblock \doi{10.1086/380393}.
\newblock {\href{https://arxiv.org/abs/astro-ph/0310158}{{arXiv:astro-ph/0310158}}}. {[astro-ph]}.

\bibitem[\protect\citeauthoryear{{Halzen} and {Hooper}}{2005}]{2005APh....23..537H}
{Halzen} F, {Hooper} D.
\newblock {High energy neutrinos from the TeV Blazar 1ES 1959 + 650}.
\newblock Astroparticle Physics. 2005 Jul;23(6):537--542.
\newblock \doi{10.1016/j.astropartphys.2005.03.007}.
\newblock {\href{https://arxiv.org/abs/astro-ph/0502449}{{arXiv:astro-ph/0502449}}}. {[astro-ph]}.

\bibitem[\protect\citeauthoryear{{Tagliaferri} et~al.}{2008}]{2008ApJ...679.1029T}
{Tagliaferri} G, {Foschini} L, {Ghisellini} G, {Maraschi} L, {pre=''and'' post=''}, {and'' affil=''4''>G  Tosti} a, et~al.
\newblock {Simultaneous Multiwavelength Observations of the Blazar 1ES 1959+650 at a Low TeV Flux}.
\newblock \apj. 2008 Jun;679(2):1029--1039.
\newblock \doi{10.1086/586731}.
\newblock {\href{https://arxiv.org/abs/0801.4029}{{arXiv:0801.4029}}}. {[astro-ph]}.

\bibitem[\protect\citeauthoryear{{Tavecchio} et~al.}{2010}]{2010MNRAS.401.1570T}
{Tavecchio} F, {Ghisellini} G, {Ghirlanda} G, {Foschini} L, {Maraschi} L.
\newblock {TeV BL Lac objects at the dawn of the Fermi era}.
\newblock \mnras. 2010 Jan;401(3):1570--1586.
\newblock \doi{10.1111/j.1365-2966.2009.15784.x}.
\newblock {\href{https://arxiv.org/abs/0909.0651}{{arXiv:0909.0651}}}. {[astro-ph.HE]}.

\bibitem[\protect\citeauthoryear{{H.~E.~S.~S. Collaboration} et~al.}{2010}]{2010A&A...511A..52H}
{H ~E ~S ~S  Collaboration}, {Acero} F, {Aharonian} F, {Akhperjanian} AG, {Anton} G, {Barres de Almeida} U, et~al.
\newblock {PKS 2005-489 at VHE: four years of monitoring with HESS and simultaneous multi-wavelength observations}.
\newblock \aap. 2010 Feb;511:A52.
\newblock \doi{10.1051/0004-6361/200913073}.
\newblock {\href{https://arxiv.org/abs/0911.2709}{{arXiv:0911.2709}}}. {[astro-ph.CO]}.

\bibitem[\protect\citeauthoryear{{Aharonian} et~al.}{2009}]{2009A&A...502..749A}
{Aharonian} F, {Akhperjanian} AG, {Anton} G, {Barres de Almeida} U, {Bazer-Bachi} AR, {Becherini} Y, et~al.
\newblock {Simultaneous multiwavelength observations of the second exceptional {\ensuremath{\gamma}}-ray flare of PKS 2155-304 in July 2006}.
\newblock \aap. 2009 Aug;502(3):749--770.
\newblock \doi{10.1051/0004-6361/200912128}.
\newblock {\href{https://arxiv.org/abs/0906.2002}{{arXiv:0906.2002}}}. {[astro-ph.CO]}.

\bibitem[\protect\citeauthoryear{{Bhagwan} et~al.}{2016}]{2016NewA...44...21B}
{Bhagwan} J, {Gupta} AC, {Papadakis} IE, {Wiita} PJ.
\newblock {Flux and spectral variability of the blazar PKS 2155 -304 with XMM-Newton: Evidence of particle acceleration and synchrotron cooling}.
\newblock \na. 2016 Apr;44:21--28.
\newblock \doi{10.1016/j.newast.2015.08.005}.
\newblock {\href{https://arxiv.org/abs/1510.02817}{{arXiv:1510.02817}}}. {[astro-ph.HE]}.

\bibitem[\protect\citeauthoryear{Consortium}{2019}]{CTACollaboration}
Consortium CTA.
\newblock {Science with the Cherenkov Telescope Array}; 2019.

\bibitem[\protect\citeauthoryear{{Tchernin} et~al.}{2013}]{2013A&A...555A..70T}
{Tchernin} C, {Aguilar} JA, {Neronov} A, {Montaruli} T.
\newblock {An exploration of hadronic interactions in blazars using IceCube}.
\newblock \aap. 2013 Jul;555:A70.
\newblock \doi{10.1051/0004-6361/201220508}.
\newblock {\href{https://arxiv.org/abs/1305.3524}{{arXiv:1305.3524}}}. {[astro-ph.HE]}.

\bibitem[\protect\citeauthoryear{{Zandanel} et~al.}{2015}]{2015A&A...578A..32Z}
{Zandanel} F, {Tamborra} I, {Gabici} S, {Ando} S.
\newblock {High-energy gamma-ray and neutrino backgrounds from clusters of galaxies and radio constraints}.
\newblock \aap. 2015 Jun;578:A32.
\newblock \doi{10.1051/0004-6361/201425249}.
\newblock {\href{https://arxiv.org/abs/1410.8697}{{arXiv:1410.8697}}}. {[astro-ph.HE]}.

\bibitem[\protect\citeauthoryear{{Sahakyan}}{2018}]{2018ApJ...866..109S}
{Sahakyan} N.
\newblock {Lepto-hadronic {\ensuremath{\gamma}}-Ray and Neutrino Emission from the Jet of TXS 0506+056}.
\newblock \apj. 2018 Oct;866(2):109.
\newblock \doi{10.3847/1538-4357/aadade}.
\newblock {\href{https://arxiv.org/abs/1808.05651}{{arXiv:1808.05651}}}. {[astro-ph.HE]}.

\bibitem[\protect\citeauthoryear{{Aguilar-Ruiz} et~al.}{2023}]{2023Galax..11..117A}
{Aguilar-Ruiz} E, {Galv{\'a}n-G{\'a}mez} A, {Fraija} N.
\newblock {Testing a Lepto-Hadronic Two-Zone Model with Extreme High-Synchrotron Peaked BL Lacs and Track-like High-Energy Neutrinos}.
\newblock Galaxies. 2023 Dec;11(6):117.
\newblock \doi{10.3390/galaxies11060117}.

\bibitem[\protect\citeauthoryear{{Aguilar-Ruiz} et~al.}{2023}]{2023JHEAp..38....1A}
{Aguilar-Ruiz} E, {Fraija} N, {Galv{\'a}n-G{\'a}mez} A.
\newblock {High-energy neutrino fluxes from hard-TeV BL Lacs}.
\newblock Journal of High Energy Astrophysics. 2023 Jun;38:1--11.
\newblock \doi{10.1016/j.jheap.2023.02.001}.
\newblock {\href{https://arxiv.org/abs/2303.13025}{{arXiv:2303.13025}}}. {[astro-ph.HE]}.

\bibitem[\protect\citeauthoryear{{Cerruti}}{2020}]{2020Galax...8...72C}
{Cerruti} M.
\newblock {Leptonic and Hadronic Radiative Processes in Supermassive-Black-Hole Jets}.
\newblock Galaxies. 2020 Oct;8(4):72.
\newblock \doi{10.3390/galaxies8040072}.
\newblock {\href{https://arxiv.org/abs/2012.13302}{{arXiv:2012.13302}}}. {[astro-ph.HE]}.

\bibitem[\protect\citeauthoryear{{Owen} and {Yang}}{2022}]{2022MNRAS.516.1539O}
{Owen} ER, {Yang} HYK.
\newblock {Emission from hadronic and leptonic processes in galactic jet-driven bubbles}.
\newblock \mnras. 2022 Oct;516(1):1539--1556.
\newblock \doi{10.1093/mnras/stac2289}.
\newblock {\href{https://arxiv.org/abs/2111.01402}{{arXiv:2111.01402}}}. {[astro-ph.HE]}.

\bibitem[\protect\citeauthoryear{{Linden} et~al.}{2012}]{2012ApJ...753...41L}
{Linden} T, {Lovegrove} E, {Profumo} S.
\newblock {The Morphology of Hadronic Emission Models for the Gamma-Ray Source at the Galactic Center}.
\newblock \apj. 2012 Jul;753(1):41.
\newblock \doi{10.1088/0004-637X/753/1/41}.
\newblock {\href{https://arxiv.org/abs/1203.3539}{{arXiv:1203.3539}}}. {[astro-ph.HE]}.

\bibitem[\protect\citeauthoryear{{MAGIC Collaboration} et~al.}{2023}]{2023A&A...671A..12M}
{MAGIC Collaboration}, {Abe} H, {Abe} S, {Acciari} VA, {Agudo} I, {Aniello} T, et~al.
\newblock {MAGIC observations provide compelling evidence of hadronic multi-TeV emission from the putative PeVatron SNR G106.3+2.7}.
\newblock \aap. 2023 Mar;671:A12.
\newblock \doi{10.1051/0004-6361/202244931}.
\newblock {\href{https://arxiv.org/abs/2211.15321}{{arXiv:2211.15321}}}. {[astro-ph.HE]}.

\bibitem[\protect\citeauthoryear{{Sol} and {Zech}}{2022}]{2022Galax..10..105S}
{Sol} H, {Zech} A.
\newblock {Blazars at Very High Energies: Emission Modelling}.
\newblock Galaxies. 2022 Nov;10(6):105.
\newblock \doi{10.3390/galaxies10060105}.
\newblock {\href{https://arxiv.org/abs/2211.03580}{{arXiv:2211.03580}}}. {[astro-ph.HE]}.

\bibitem[\protect\citeauthoryear{{Zhang} et~al.}{2024}]{2024ApJ...967...93Z}
{Zhang} H, {B{\"o}ttcher} M, {Liodakis} I.
\newblock {Revisiting High-energy Polarization from Leptonic and Hadronic Blazar Scenarios}.
\newblock \apj. 2024 Jun;967(2):93.
\newblock \doi{10.3847/1538-4357/ad4112}.
\newblock {\href{https://arxiv.org/abs/2404.12475}{{arXiv:2404.12475}}}. {[astro-ph.HE]}.

\bibitem[\protect\citeauthoryear{Postnikov et~al.}{2017}]{Postnikov2017}
Postnikov EB, Grinyuk AA, Kuzmichev LA, Sveshnikova LG.
\newblock Hybrid method for identifying mass groups of primary cosmic rays in the joint operation of IACTs and wide angle Cherenkov timing arrays.
\newblock Journal of Physics: Conference Series. 2017 jan;798(1):012030.
\newblock \doi{10.1088/1742-6596/798/1/012030}.

\bibitem[\protect\citeauthoryear{{Wagner}}{2022}]{Wagner2022}
{Wagner} S.
\newblock {Observations of Cosmic Ray sources with the HESS IACT-array and new results on cosmic ray propagation}.
\newblock In: 44th COSPAR Scientific Assembly. Held 16-24 July. vol.~44; 2022. p. 2144.

\bibitem[\protect\citeauthoryear{Acero et~al.}{2023}]{ACERO2023}
Acero F, Acharyya A, Adam R, Aguasca-Cabot A, Agudo I, Aguirre-Santaella A, et~al.
\newblock Sensitivity of the Cherenkov Telescope Array to spectral signatures of hadronic PeVatrons with application to Galactic Supernova Remnants.
\newblock Astroparticle Physics. 2023;150:102850.
\newblock \doi{https://doi.org/10.1016/j.astropartphys.2023.102850}.

\bibitem[\protect\citeauthoryear{{Bell}}{2013}]{2013APh....43...56B}
{Bell} AR.
\newblock {Cosmic ray acceleration}.
\newblock Astroparticle Physics. 2013 Mar;43:56--70.
\newblock \doi{10.1016/j.astropartphys.2012.05.022}.

\bibitem[\protect\citeauthoryear{}{}]{MAGIC}
: The MAGIC gamma-ray observatory.
\newblock Accessed: 2023-09-15.
\newblock \url{https://wwwmagic.mppmu.mpg.de}.

\bibitem[\protect\citeauthoryear{}{}]{Veritas}
: VERITAS - Very Energetic Radiation Imaging Telescope Array System.
\newblock Accessed: 2023-09-15.
\newblock \url{https://veritas.sao.arizona.edu}.

\bibitem[\protect\citeauthoryear{}{}]{HESS}
: H.E.S.S. - The High Energy Stereoscopic System.
\newblock Accessed: 2023-09-15.
\newblock \url{https://www.mpi-hd.mpg.de/hfm/HESS/}.

\bibitem[\protect\citeauthoryear{{CTA Consortium} and {Ong}}{2019}]{2019EPJWC.20901038C}
{CTA Consortium}, {Ong} RA.
\newblock {The Cherenkov Telescope Array Science Goals and Current Status}.
\newblock In: European Physical Journal Web of Conferences. vol. 209 of European Physical Journal Web of Conferences; 2019. p. 01038.

\bibitem[\protect\citeauthoryear{{di Sciascio} and {Lhaaso Collaboration}}{2016}]{2016NPPP..279..166D}
{di Sciascio} G, {Lhaaso Collaboration}.
\newblock {The LHAASO experiment: From Gamma-Ray Astronomy to Cosmic Rays}.
\newblock Nuclear and Particle Physics Proceedings. 2016 Oct;279-281:166--173.
\newblock \doi{10.1016/j.nuclphysbps.2016.10.024}.
\newblock {\href{https://arxiv.org/abs/1602.07600}{{arXiv:1602.07600}}}. {[astro-ph.HE]}.

\bibitem[\protect\citeauthoryear{{Cao} et~al.}{2023}]{2023PhRvL.131o1001C}
{Cao} Z, {Aharonian} F, {An} Q, {Axikegu} YX Bai, {Bao} YW, {Bastieri} D, et~al.
\newblock {Measurement of Ultra-High-Energy Diffuse Gamma-Ray Emission of the Galactic Plane from 10 TeV to 1 PeV with LHAASO-KM2A}.
\newblock \prl. 2023 Oct;131(15):151001.
\newblock \doi{10.1103/PhysRevLett.131.151001}.
\newblock {\href{https://arxiv.org/abs/2305.05372}{{arXiv:2305.05372}}}. {[astro-ph.HE]}.

\bibitem[\protect\citeauthoryear{{Chiavassa} and {SWGO Collaboration}}{2024}]{2024JInst..19C2065C}
{Chiavassa} A, {SWGO Collaboration}.
\newblock {SWGO: a wide-field of view gamma-ray observatory in the southern hemisphere}.
\newblock Journal of Instrumentation. 2024 Feb;19(2):C02065.
\newblock \doi{10.1088/1748-0221/19/02/C02065}.

\bibitem[\protect\citeauthoryear{{Hu} et~al.}{2024}]{2024MNRAS.531.5061H}
{Hu} Q, {Zhang} Y, {Duan} K, {Zeng} H.
\newblock {Prospects for the detection of very-high-energy pulsars with LHAASO and SWGO}.
\newblock \mnras. 2024 Jul;531(4):5061--5066.
\newblock \doi{10.1093/mnras/stae1497}.
\newblock {\href{https://arxiv.org/abs/2407.00262}{{arXiv:2407.00262}}}. {[astro-ph.HE]}.

\bibitem[\protect\citeauthoryear{Klinger et~al.}{2024}]{Klinger_2024}
Klinger M, Rudolph A, Rodrigues X, Yuan C, de~Clairfontaine GF, Fedynitch A, et~al.
\newblock AM3: An Open-source Tool for Time-dependent Lepto-hadronic Modeling of Astrophysical Sources.
\newblock The Astrophysical Journal Supplement Series. 2024 oct;275(1):4.
\newblock \doi{10.3847/1538-4365/ad725c}.

\bibitem[\protect\citeauthoryear{{Klinger} et~al.}{2023}]{AM3}
{Klinger} M, {Rudolph} A, {Rodrigues} X, {Yuan} C, {Fichet de Clairfontaine} G, {Fedynitch} A, et~al.
\newblock {AM$^3$: An Open-Source Tool for Time-Dependent Lepto-Hadronic Modeling of Astrophysical Sources}.
\newblock arXiv e-prints. 2023 Dec;p. arXiv:2312.13371.
\newblock \doi{10.48550/arXiv.2312.13371}.
\newblock {\href{https://arxiv.org/abs/2312.13371}{{arXiv:2312.13371}}}. {[astro-ph.HE]}.

\bibitem[\protect\citeauthoryear{{Ostapchenko}}{2024}]{2024PhRvD.109c4002O}
{Ostapchenko} S.
\newblock {QGSJET-III model of high energy hadronic interactions: The formalism}.
\newblock \prd. 2024 Feb;109(3):034002.
\newblock \doi{10.1103/PhysRevD.109.034002}.
\newblock {\href{https://arxiv.org/abs/2401.06202}{{arXiv:2401.06202}}}. {[hep-ph]}.

\bibitem[\protect\citeauthoryear{Acciari et~al.}{2008}]{Acciari_2008}
Acciari VA, Aliu E, Beilicke M, Benbow W, Böttcher M, Bradbury SM, et~al.
\newblock VERITAS Discovery of $>$ 200 GeV Gamma-Ray Emission from the Intermediate-Frequency-Peaked BL Lacertae Object W Comae.
\newblock The Astrophysical Journal. 2008 aug;684(2):L73.
\newblock \doi{10.1086/592244}.

\bibitem[\protect\citeauthoryear{Biteau and Williams}{2015}]{Biteau_2015}
Biteau J, Williams DA.
\newblock THE EXTRAGALACTIC BACKGROUND LIGHT, THE HUBBLE CONSTANT, AND ANOMALIES: CONCLUSIONS FROM 20 YEARS OF TeV GAMMA-RAY OBSERVATIONS.
\newblock The Astrophysical Journal. 2015 oct;812(1):60.
\newblock \doi{10.1088/0004-637X/812/1/60}.

\bibitem[\protect\citeauthoryear{{HESS Collaboration} et~al.}{2010}]{HessPKS2005}
{HESS Collaboration}, {Acero, F }, {Aharonian, F }, {Akhperjanian, A  G }, {Anton, G }, {Barres de Almeida, U }, et~al.
\newblock PKS2005-489 at VHE: four years of monitoring with HESS and simultaneous multi-wavelength observations******.
\newblock A\&A. 2010;511:A52.
\newblock \doi{10.1051/0004-6361/200913073}.

\bibitem[\protect\citeauthoryear{Collaboration}{2023}]{2023ApJ...954...75A}
Collaboration I.
\newblock {Search for Correlations of High-energy Neutrinos Detected in IceCube with Radio-bright AGN and Gamma-Ray Emission from Blazars}.
\newblock \apj. 2023 Sep;954(1):75.
\newblock \doi{10.3847/1538-4357/acdfcb}.
\newblock {\href{https://arxiv.org/abs/2304.12675}{{arXiv:2304.12675}}}. {[astro-ph.HE]}.

\bibitem[\protect\citeauthoryear{Costa et~al.}{2024}]{Costa2024}
Costa RJ, Götz DB, Anjos RC, Pereira LAS, Mello AJTS.
\newblock A gamma-ray study of galactic PeVatron candidates LHAASO J1825-1326 and LHAASO J1839-0545.
\newblock Journal of Cosmology and Astroparticle Physics. 2024 7;2024:035.
\newblock \doi{10.1088/1475-7516/2024/07/035}.

\bibitem[\protect\citeauthoryear{{Donath, Axel} et~al.}{2023}]{Donath_2023}
{Donath, Axel}, {Terrier, Régis}, {Remy, Quentin}, {Sinha, Atreyee}, {Nigro, Cosimo}, {Pintore, Fabio}, et~al.
\newblock Gammapy: A Python package for gamma-ray astronomy.
\newblock A\&A. 2023;678:A157.
\newblock \doi{10.1051/0004-6361/202346488}.

\bibitem[\protect\citeauthoryear{Deil et~al.}{2017}]{Deil2017}
Deil C, Zanin R, Lefaucheur J, Boisson C, Khélifi B, Terrier R, et~al.
\newblock Gammapy - A prototype for the CTA science tools; 2017. .

\bibitem[\protect\citeauthoryear{Acharyya et~al.}{2023}]{Acharyya_2023}
Acharyya A, Adams CB, Archer A, Bangale P, Bartkoske JT, Batista P, et~al.
\newblock VTSCat: The VERITAS Catalog of Gamma-Ray Observations.
\newblock Research Notes of the AAS. 2023 jan;7(1):6.
\newblock \doi{10.3847/2515-5172/acb147}.

\bibitem[\protect\citeauthoryear{Acero et~al.}{2015}]{Acero_2015}
Acero F, Ackermann M, Ajello M, Albert A, Atwood WB, Axelsson M, et~al.
\newblock Fermi Large Area Telescope Third Source Catalog.
\newblock Astrophysical Journal, Supplement Series. 2015;218.
\newblock \doi{10.1088/0067-0049/218/2/23}.

\bibitem[\protect\citeauthoryear{Abdollahi et~al.}{2022}]{Abdollahi2022}
Abdollahi S, Acero F, Baldini L, Ballet J, Bastieri D, Bellazzini R, et~al.
\newblock Incremental Fermi Large Area Telescope Fourth Source Catalog.
\newblock The Astrophysical Journal Supplement Series. 2022;260.
\newblock \doi{10.3847/1538-4365/ac6751}.

\bibitem[\protect\citeauthoryear{Ackermann et~al.}{2016}]{Ackermann_2016}
Ackermann M, Ajello M, Atwood WB, Baldini L, Ballet J, Barbiellini G, et~al.
\newblock 2FHL: THE SECOND CATALOG OF HARD FERMI-LAT SOURCES.
\newblock The Astrophysical Journal Supplement Series. 2016 1;222:5.
\newblock \doi{10.3847/0067-0049/222/1/5}.

\bibitem[\protect\citeauthoryear{Ajello et~al.}{2017}]{Ajello2017}
Ajello M, Atwood WB, Baldini L, Ballet J, Barbiellini G, Bastieri D, et~al.
\newblock 3FHL: The Third Catalog of Hard Fermi -LAT Sources.
\newblock The Astrophysical Journal Supplement Series. 2017;232.
\newblock \doi{10.3847/1538-4365/aa8221}.

\bibitem[\protect\citeauthoryear{Ballet et~al.}{2023}]{ballet2023fermi}
Ballet J, Bruel P, Burnett TH, Lott B, collaboration TFL.: Fermi Large Area Telescope Fourth Source Catalog Data Release 4 (4FGL-DR4).

\bibitem[\protect\citeauthoryear{Truebenbach and Darling}{2017}]{Truebenbach_2017}
Truebenbach AE, Darling J.
\newblock The VLBA Extragalactic Proper Motion Catalog and a Measurement of the Secular Aberration Drift.
\newblock The Astrophysical Journal Supplement Series. 2017 nov;233(1):3.
\newblock \doi{10.3847/1538-4365/aa9026}.

\bibitem[\protect\citeauthoryear{Hunt et~al.}{2021}]{Hunt_2021}
Hunt LR, Johnson MC, Cigan PJ, Gordon D, Spitzak J.
\newblock Imaging Sources in the Third Realization of the International Celestial Reference Frame.
\newblock The Astronomical Journal. 2021 aug;162(3):121.
\newblock \doi{10.3847/1538-3881/ac135d}.

\bibitem[\protect\citeauthoryear{Xu et~al.}{2019}]{Xu_2019}
Xu MH, Anderson JM, Heinkelmann R, Lunz S, Schuh H, Wang GL.
\newblock Structure Effects for 3417 Celestial Reference Frame Radio Sources.
\newblock The Astrophysical Journal Supplement Series. 2019 may;242(1):5.
\newblock \doi{10.3847/1538-4365/ab16ea}.

\bibitem[\protect\citeauthoryear{Bail et~al.}{2016}]{LeBail_2016}
Bail KL, Gipson JM, Gordon D, MacMillan DS, Behrend D, Thomas CC, et~al.
\newblock IVS OBSERVATION OF ICRF2-GAIA TRANSFER SOURCES.
\newblock The Astronomical Journal. 2016 feb;151(3):79.
\newblock \doi{10.3847/0004-6256/151/3/79}.

\bibitem[\protect\citeauthoryear{Saldana-Lopez et~al.}{2021}]{Saldana_Lopez_2021}
Saldana-Lopez A, Domínguez A, Pérez-González PG, Finke J, Ajello M, Primack JR, et~al.
\newblock An observational determination of the evolving extragalactic background light from the multiwavelength HST/CANDELS survey in the Fermi and CTA era.
\newblock Monthly Notices of the Royal Astronomical Society. 2021 Aug;507(4):5144–5160.
\newblock \doi{10.1093/mnras/stab2393}.

\bibitem[\protect\citeauthoryear{Akaike}{1974}]{Akaike1974}
Akaike H.
\newblock A new look at the statistical model identification.
\newblock IEEE Transactions on Automatic Control. 1974;19(6):716--723.
\newblock \doi{10.1109/TAC.1974.1100705}.

\bibitem[\protect\citeauthoryear{{Di Piano, Ambra and Bulgarelli, Andrea and Fioretti, Valentina and Baroncelli, Leonardo and Parmiggiani, Nicol\`o and Longo, Francesco and Stamerra, Antonio and L\'opez-Oramas, Alicia and Stratta, Giulia and De Cesare, and Giovanni}}{2021}]{Piano2021}
{Di Piano, Ambra and Bulgarelli, Andrea and Fioretti, Valentina and Baroncelli, Leonardo and Parmiggiani, Nicol\`o and Longo, Francesco and Stamerra, Antonio and L\'opez-Oramas, Alicia and Stratta, Giulia and De Cesare, and Giovanni}.
\newblock {Detection methods for the Cherenkov Telescope Array at very-short exposure times}.
\newblock PoS. 2021;ICRC2021:694.
\newblock \doi{10.22323/1.395.0694}.
\newblock {\href{https://arxiv.org/abs/2108.04504}{{arXiv:2108.04504}}}. {[astro-ph.IM]}.

\bibitem[\protect\citeauthoryear{Observatory and Consortium}{2021}]{CTAOIRFS}
Observatory CTA, Consortium CTA.: {CTAO Instrument Response Functions - prod5 version v0.1}.
\newblock Zenodo.
\newblock Available from: \url{https://doi.org/10.5281/zenodo.5499840}.

\bibitem[\protect\citeauthoryear{Manchester et~al.}{2005}]{Manchester2005}
Manchester RN, Hobbs GB, Teoh A, Hobbs M.
\newblock The Australia Telescope National Facility Pulsar Catalogue.
\newblock The Astronomical Journal. 2005 4;129:1993--2006.
\newblock \doi{10.1086/428488}.

\end{thebibliography}

\end{document}